\documentclass[a4paper,10pt,twoside]{cpc-hepnp}
\usepackage{CJK,upgreek,fancyhdr}
\usepackage{multicol}
\usepackage{graphicx}
\usepackage{booktabs}
\usepackage{amssymb,bm,mathrsfs,bbm,amscd}
\usepackage[tbtags]{amsmath}
\usepackage{lastpage}
\usepackage{epstopdf}
\usepackage[colorlinks, allcolors=blue]{hyperref}
\usepackage{array}

\begin{document}
\begin{CJK*}{GB}{gbsn}

\fancyhead[c]{\small Chinese Physics C Vol. 41, No. 7 (2017) 077001\quad(DOI: 10.1088/1674-1137/41/7/077001)}
\fancyfoot[C]{\small 077001-\thepage}

\footnotetext[0]{Received 16 February 2017}

\title{Beam distribution reconstruction simulation for electron beam probe}

\author{%
       Yong-Chun Feng$^{1,2}$
\quad Rui-Shi Mao$^{1;1)}$\email{maorsh@impcas.ac.cn}%
\quad Peng Li$^{1}$
\quad Xin-Cai Kang$^{1}$
\\
\quad Yan Yin$^{1}$
\quad Tong Liu$^{1,2}$
\quad Yao-Yao You$^{1,2}$
\quad Yu-Cong Chen$^{1}$
\\
\quad Tie-Cheng Zhao$^{1}$
\quad Zhi-Guo Xu$^{1}$
\quad Yan-Yu Wang$^{1}$
\quad You-Jin Yuan$^{1}$
}
\maketitle

\address{%
$^1$ Institute of Modern Physics, Chinese Academy of Sciences, Lanzhou 730000, China\\
$^2$ University of Chinese Academy of Sciences, Beijing 100049, China\\
}

\begin{abstract}
An electron beam probe (EBP) is a detector which makes use of a low-intensity and low-energy electron beam to measure the transverse profile, bunch shape, beam neutralization and beam wake field of an intense beam with small dimensions. While it can be applied to many aspects, we limit our analysis to beam distribution reconstruction. This kind of detector is almost non-interceptive for all of the beam and does not disturb the machine environment. In this paper, we present the theoretical aspects behind this technique for beam distribution measurement and some simulation results of the detector involved. First, a method to obtain a parallel electron beam is introduced and a simulation code is developed. An EBP as a profile monitor for dense beams is then simulated using the fast scan method for various target beam profiles, including KV distribution, waterbag distribution, parabolic distribution, Gaussian distribution and halo distribution. Profile reconstruction from the deflected electron beam trajectory is implemented and compared with the actual profile, and the expected agreement is achieved. Furthermore, as well as fast scan, a slow scan, i.e. step-by-step scan, is considered, which lowers the requirement for hardware, i.e. Radio Frequency deflector. We calculate the three-dimensional electric field of a Gaussian distribution and simulate the electron motion in this field. In addition, a fast scan along the target beam direction and slow scan across the beam are also presented, and can provide a measurement of longitudinal distribution as well as transverse profile simultaneously. As an example, simulation results for the China Accelerator Driven Sub-critical System (CADS) and High Intensity Heavy Ion Accelerator Facility (HIAF) are given. Finally, a potential system design for an EBP is described.
\end{abstract}

\begin{keyword}
reconstruction, beam distribution, electron beam probe, simulation
\end{keyword}

\begin{pacs}
29.20.db, 29.27.Bd
\end{pacs}

\footnotetext[0]{\hspace*{-3mm}\raisebox{0.3ex}{$\scriptstyle\copyright$}2017
Chinese Physical Society and the Institute of High Energy Physics
of the Chinese Academy of Sciences and the Institute
of Modern Physics of the Chinese Academy of Sciences and IOP Publishing Ltd}%

\begin{multicols}{2}

\section{\label{intro}INTRODUCTION}
Beam profile measurement is of prime importance for all accelerators, especially for high intensity machines, as it can reveal the beam width in  locations with small aperture and further match phase space between different parts of an accelerator facility. Conventional techniques~\cite{bib1} for measuring beam distribution involve a very large variety of devices depending on the beam particles, intensity and energy, such as scintillator screens, secondary electron emission grids and wire scanners. These devices typically need to insert a physical object into the beam path. Such an object is easily destroyed under increasing beam intensity and in turn can result in beam loss. So, some kinds of non-interceptive profile monitors have been launched based on different principles, such as the ionization profile monitor~\cite{bib2} (IPM), beam induced fluorescence monitor~\cite{bib3} (BIF) and electron beam probe (EBP).

The application of charged particles as a probe beam to determine charge distribution, and thus the beam profile, has been raised since the 1970s~\cite{bib4,bib5}. In those days, electron beams were used to diagnose plasma charge distribution. The development of this idea prompted accelerator scientists to study the potential of this emerging technique as an alternative approach to practically non-invasive profile monitors for accelerator beams, especially for high intensity beam. Since then, many laboratories worldwide have studied and improved EBPs and obtained some very valuable results. Among these labs, TRIUMF~\cite{bib6},  Lawrence Berkeley National Laboratory (LBNL)~cite{bib7}, the US Spallation Neutron Source (SNS)~\cite{bib8,bib9,bib10} and Fermi National Accelerator Laboratory (FNAL)~\cite{bib11,bib12} use electron beams as a probe to detect ion beam profiles. The Budker Institute of Nuclear Physics (BINP)~\cite{bib13,bib14,bib15,bib16} in Russia uses electron beams to measure ultra-relativistic electron bunch length and beam distribution. In addition, some labs, e.g. the European Organization for Nuclear Research (CERN)~\cite{bib17}, employ ion beams as a probe to extract ion beam profiles, and even some use ion beams to detect electron beams, e.g. Stanford Linear Accelerator Center (SLAC)~\cite{bib18}.

The principle behind EBPs is that a low energy, low current electron beam is injected across the target beam perpendicularly and then deflected by the target beam collective field (mainly electric field). A screen and CCD located downstream capture the deflected electron beam trace, and then, by some mathematical treatment, i.e. derivative, the beam profile can be reconstructed accurately. Since the measurement should not significantly disturb the field generated by the target beam, the current of the electron beam must be low compared to the target beam.

EBPs are suitable for both circular and linear accelerators and mainly used for high intensity beams. Two next-generation accelerator facilities, the High Intensity Heavy Ion Accelerator Facility (HIAF)~\cite{bib19} and the China Accelerator Driven Sub-critical System (CADS)~\cite{bib20}, have been proposed by the Institute of Modern Physics (IMP).
Both systems have high intensity or high power, energy 1.2~GeV/u with intensity 5 $\times 10^{11}$ ppp (particles per pulse) for HIAF and 10 MW for CADS phase \uppercase\expandafter{\romannumeral1}. Measuring the beam parameters of these high power accelerators is challenging and usually depends on non-invasive instruments. An EBP detector may provide the capacity for fine accelerator tuning and online control of beam stability.

The purpose of this article is to present the study and simulation of an EBP under various target beam profile distributions. Some interesting results have been achieved with fast scan and slow scan. An example is also presented to deepen the understanding of the EBP. This simulation is expected to provide the theoretical basis for testing and construction of an EBP in the future.

\section{PRINCIPLE}
An EBP uses the deflection of a low energy probe beam in the target beam electromagnetic field to infer the profile information of the target beam. Measuring the deflection angle as a function of different impacts, one can reconstruct the beam distribution in the $x$ or $y$ direction. The theory of profile reconstruction and the validity of this theory are presented below.
\subsection{Theory of profile reconstruction}
Without loss of generality, assume electron beam has a tilted incident angle~\cite{bib21} with an impact parameter $\rho$, as depicted in Fig. \ref{fig1}. The target beam moves along the $z$ direction, centered at $x=y=0$. Neglecting magnetic field, the transverse electric field can be divided into perpendicular and parallel components.
\begin{equation}
E_{\perp}=E_x cos\phi + E_y sin\phi, \quad E_{\parallel}=E_x sin\phi - E_y cos\phi
\label{eq1}
\end{equation}
where $E_x$ and $E_y$ are the horizontal and vertical components of the target beam space charge electric field respectively. $\phi$ is the angle between the impact parameter and $x$ direction, and $\phi=0$ means the incident direction is $x$, $\phi=\frac{\pi}{2}$ for $y$. In the case of steady beam current, according to the Maxwell-Faraday equation, electric field is curl free.
\begin{equation}
\nabla \times \vec E=0
\label{eq2}
\end{equation}
Assume electron beam is injected at $-x_0$ and ended at $x_0$. Then, Eq.(\ref{eq2}) becomes
\begin{equation}
\int_{-x_0}^{x_0} E_\parallel d\parallel = 0
\label{eq3}
\end{equation}
The net energy change along the electron trajectory is zero if we use the above-mentioned hypothesis. Hence, we can assume the electron beam has a constant velocity $v$, which is important for this theory and is also reasonable in some sense (see Section~\ref{B}).
\begin{center}
\includegraphics[width=8.5cm]{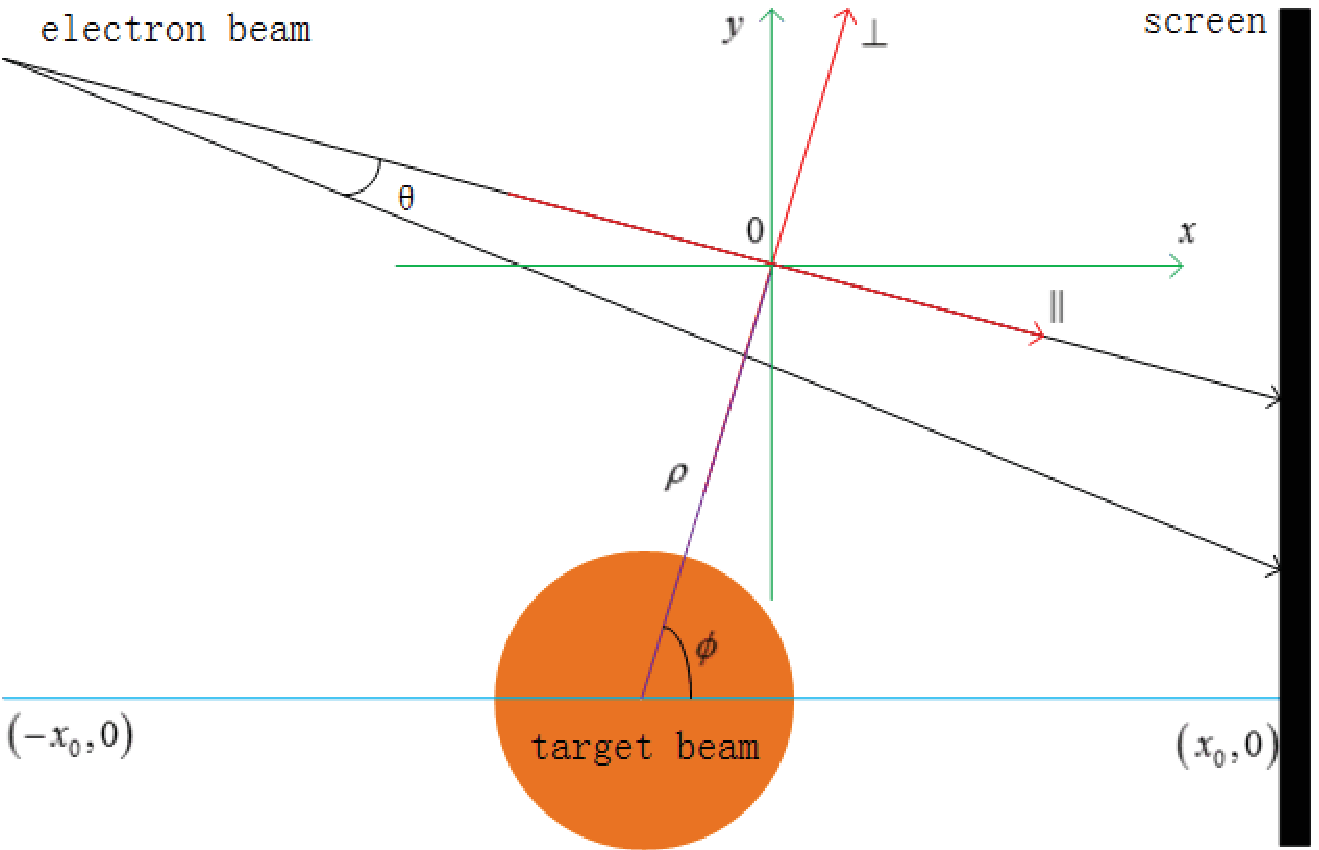}
\figcaption{\label{fig1}Schematic of electron beam deflection by the target beam.}
\end{center}

Next, we investigate the perpendicular direction. Combining Newton's second law of motion and some simple mathematical treatment, we obtain
\begin{equation}
\frac{d\perp}{d\parallel}=\frac{e}{mv^2}\int (E_x cos\phi + E_y sin\phi)d\parallel
\label{eq4}
\end{equation}
where $e$ and $m$ are the electron charge and mass, respectively. For small deflection angles,
\begin{equation}
\theta=\frac{e}{mv^2}\int (E_x cos\phi + E_y sin\phi)d\parallel
\label{eq5}
\end{equation}
Letting $\phi=\frac{\pi}{2}$, we can obtain the deflection angle along the y direction.
\begin{equation}
\theta_y=\frac{e}{mv^2}\int E_y dx
\label{eq6}
\end{equation}
After differentiating,
\begin{equation}
\frac{d\theta_y}{dy}=\frac{e}{mv^2}\int \frac{dE_y}{dy} dx
\label{eq7}
\end{equation}
Using Gauss's law $\nabla \vec E=\frac{\sigma(x,y)}{\epsilon_0}$, we obtain
\begin{equation}
\frac{d\theta_y}{dy}=\frac{e}{mv^2}\left(\int \frac{\sigma(x,y)}{\epsilon_0} dx - \frac{d}{dx}\int E_x dx\right)
\label{eq8}
\end{equation}
Using Eq. (\ref{eq3}),
\begin{equation}
\frac{d\theta_y}{dy}=\frac{e}{\epsilon_0 mv^2} \int \sigma(x,y) dx
\label{eq9}
\end{equation}
where $\int \sigma(x,y) dx$ is the profile of the $y$ direction. The above formula states that the derivative of the probe beam deflection angle with respect to impact parameter gives the projection profile of the beam cross-section distribution in the $y$ direction, which does the same thing as the wire scanner did.
\subsection{\label{B}Validity of theory}
In deriving the profile reconstruction procedure above, we introduced three important hypotheses. Firstly, we neglect the magnetic field of the target beam, because the magnetic field is around the $\phi$ direction, which has no influence on the deflection angle along the $y$ direction. In addition, for a non-relativistic beam, the magnetic field is much smaller than the electric field. We further consider that the electron beam velocity remains constant throughout the scan. However, since the electric field component along the $x$ direction exerts a force on the electron, the electron beam velocity will change slightly during its passage, although the net energy change is always zero according to symmetry. To get rid of the error due to velocity change as much as possible, the electron beam energy should be much higher than the target beam potential. We also assume that the deflection angle is small, which can be achieved with high electron beam  energy. Errors will inevitably be introduced due to each of these  hypotheses. Therefore, computer simulation is urgently needed.

\section{PRODUCING PARALLEL ELECTRON BEAM}
\begin{figure*}[htb]
\centering
\includegraphics[width=15cm]{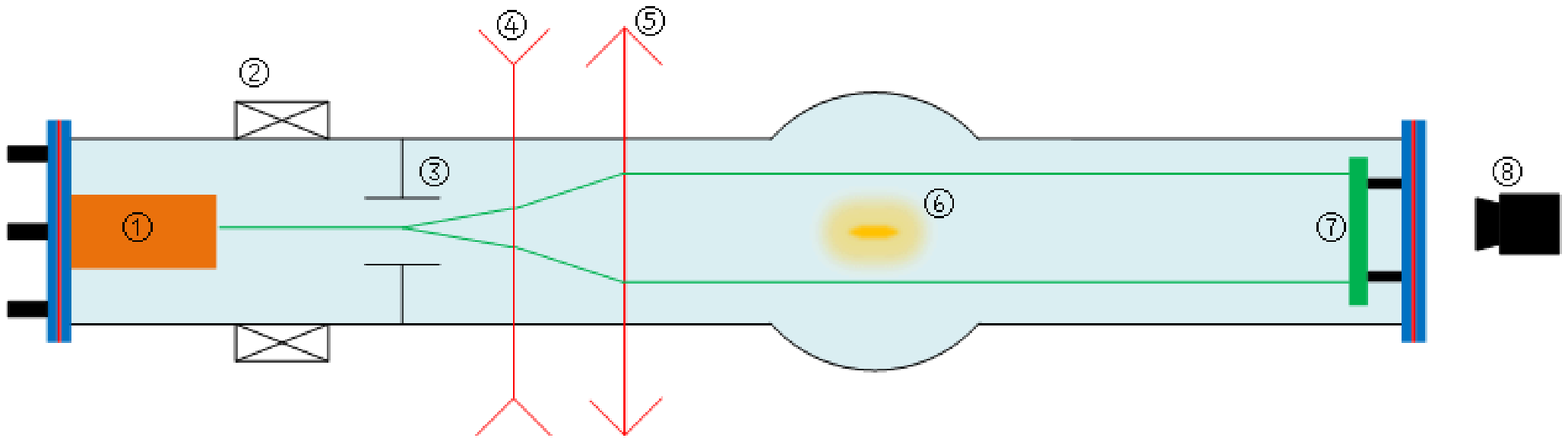}
\caption{\label{fig2}Layout of electron beam probe: 1. electron gun; 2. solenoid; 3. Radio Frequency deflector; 4. defocusing quadrupole; 5. focusing quadrupole; 6. target beam; 7. YaG:Ce screen; 8. CCD.}
\end{figure*}

To obtain the profile of the $y$ or $x$ direction, the electron beam should be scanned along $x$ or $y$ with varying $y$ or $x$ values. The key point to reconstruct beam profile is that electron beam has to be parallel to either axis and perpendicular to the target beam. In general, there are two ways of producing parallel electron beams. Lawrence Berkeley National Laboratory use four dipole magnets of equal strengths to form a chicane system~\cite{bib7}, which is similar to a bump system. By virtue of this arrangement, the electron beam can be swept in the $y$ axis while remaining parallel to the $x$ axis in the gap between the middle two magnets. Although the technology has been successfully applied and ion beam profiles  reconstructed, it cannot be used for fast scans and thus the profile cannot be measured automatically and rapidly. This situation will change with the development of another method to produce parallel electron beam, which is shown in Fig. \ref{fig2}. It is an advanced configuration used by several labs, such as SNS, BINP and FNAL.

This system consists of an electron gun for electron beam generation,  solenoid for electron focusing, RF deflector for fast scan, two thin quadrupoles for forming parallel electron beam, and an optical image system. To separate the deflected and undeflected trajectories, electrons are scanned through the target beam at a tilted angle, i.e. 45 degrees. If the scan is aligned vertically, one has to analyze the density distribution of the projected electron beam, and through simulations, this gives poor quality results~\cite{bib10}.

The mathematical model~\cite{bib22} to simulate a parallel electron beam is presented below.
For simplicity, we regard the electron beam as a point charge with no transverse momentum. We assume that electrons start off at the center of the RF deflector with initial phase space coordinates at the $y$ axis,
\begin{equation}
\begin{pmatrix}
y_0\\y_0^\prime
\end{pmatrix}
=
\begin{pmatrix}
0\\\frac{U}{2Vd}x_0
\end{pmatrix}
\label{eq10}
\end{equation}
where $U=200$~V is the voltage of the RF deflector, which can be adjusted from $0$ to maximum to obtain various initial angles. $V$ is the high voltage of the electron gun cathode, which can be changed from 1~kV to 20~kV according to different target beam intensities. For high beam current, $V$ should be large in order to keep the electron beam in the screen area. $d=4$~mm is the RF deflector gap and $x_0=4$~cm the deflector length.

To observe the focusing behaviour at the $z$ axis, we let the initial phase space coordinates of the $z$ direction be
\begin{equation}
\begin{pmatrix}
z_0\\z_0^\prime
\end{pmatrix}
=
\begin{pmatrix}
0\\0.05
\end{pmatrix}
\label{eq11}
\end{equation}
The transfer matrix along the transfer line is given by
\begin{equation}
M_y
=
\begin{pmatrix}
1&l_3\\
0&1
\end{pmatrix}
\begin{pmatrix}
1&0\\K_2&1
\end{pmatrix}
\begin{pmatrix}
1&l_2\\0&1
\end{pmatrix}
\begin{pmatrix}
1&0\\K_1&1
\end{pmatrix}
\begin{pmatrix}
1&l_1\\0&1
\end{pmatrix}
\label{eq12}
\end{equation}
\begin{equation}
M_z
=
\begin{pmatrix}
1&l_3\\0&1
\end{pmatrix}
\begin{pmatrix}
1&0\\-K_2&1
\end{pmatrix}
\begin{pmatrix}
1&l_2\\0&1
\end{pmatrix}
\begin{pmatrix}
1&0\\-K_1&1
\end{pmatrix}
\begin{pmatrix}
1&l_1\\0&1
\end{pmatrix}
\label{eq13}
\end{equation}
where $K_1$ and $K_2$ are the integrated gradients of the first and second quadrupoles respectively, $l_3$ is the distance from the focusing quadrupole center to the screen, $l_2$ is the distance from the defocusing quadrupole center to the focusing quadrupole center, and $l_1$ is the distance from the center of the RF deflector to the defocusing quadrupole center. According to the condition of point to parallel transport in the $xy$ plane and point to point transport in the $xz$ plane, we have
\begin{equation}
M_y(22)=K_1l_1(1+K_2l_2)+K_2(l_2+l_1)+1=0
\label{eq14}
\end{equation}
\begin{equation}
\begin{aligned}
M_z(12)=&-K_2(l_1l_3+l_2l_3)-K_1(l_1l_3+l_2l_2)\\
 &+K_1K_2l_1l_2l_3+(l_1+l_2+l_3)=0
\end{aligned}
\label{eq15}
\end{equation}
\begin{center}
\includegraphics[width=8.5cm]{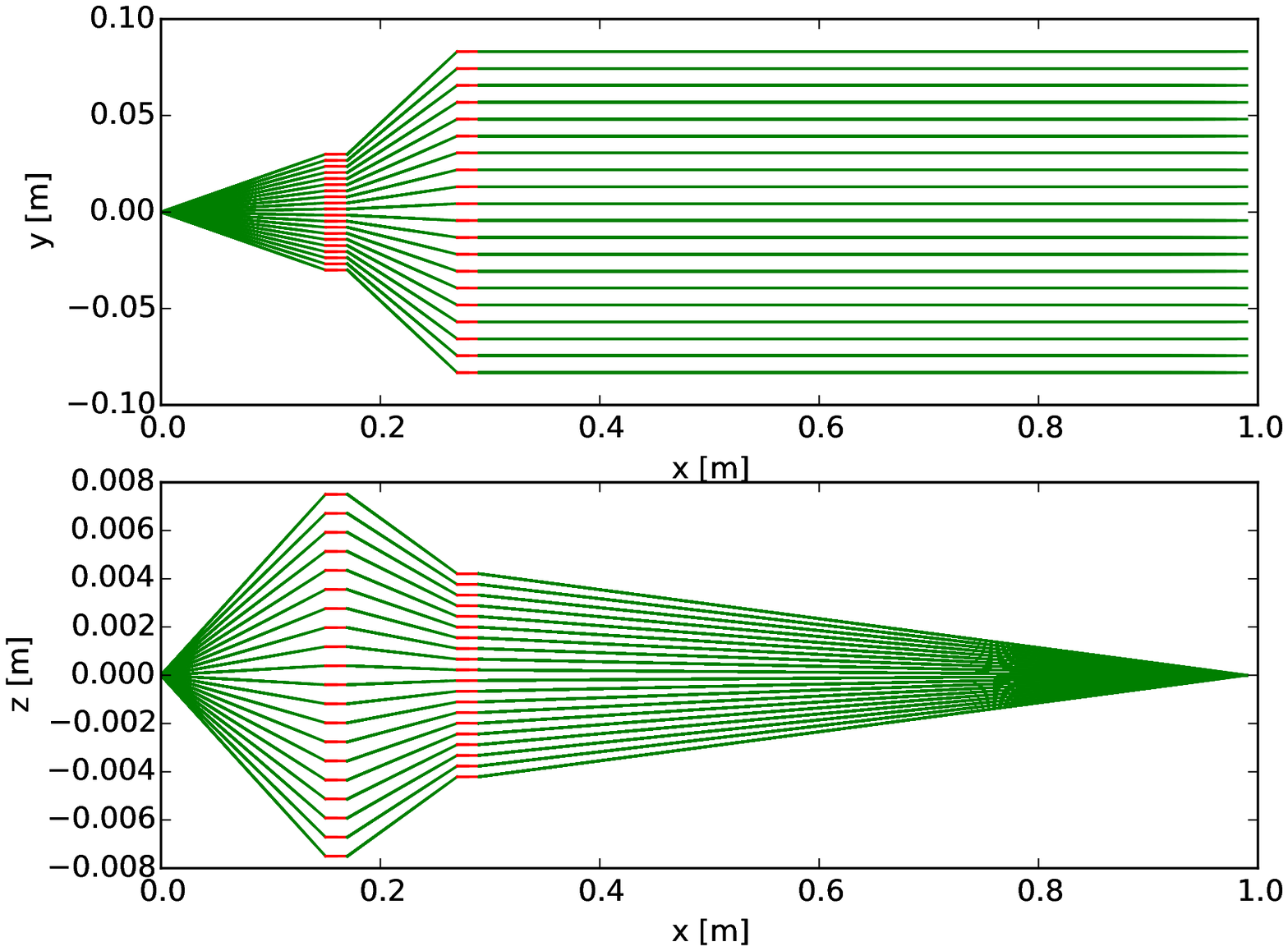}
\figcaption{\label{fig3}Top: electron trajectory in the $xy$ plane with the first solution. Bottom: electron trajectory in the $xz$ plane with the first solution.}
\end{center}
From Eq.~(\ref{eq14}) and Eq.~(\ref{eq15}), $K_1$ and $K_2$ can be expressed as an explicit function of $l_1$, $l_2$ and $l_3$. For a quadratic equation,  there exists exactly two sets of solutions,
\begin{equation}
\left\{
\begin{aligned}
K_1 &= \frac{\sqrt{l_1+l_2}\sqrt{l_1+l_2+2l_3}}{l_1\sqrt{l_2(l_2+2l_3)}} \\
K_2 &= \frac{l_1l_2+l_2^2-\sqrt{l_2(l_1+l_2)(l_2+2l_3)(l_1+l_2+2l_3)}}{2l_2(l_2+l_2)l_3}
\end{aligned}
\right.
\label{eq16}
\end{equation}
and
\begin{equation}
\left\{
\begin{aligned}
K_1 &= -\frac{\sqrt{l_1+l_2}\sqrt{l_1+l_2+2l_3}}{l_1\sqrt{l_2(l_2+2l_3)}} \\
K_2 &= \frac{l_1l_2+l_2^2+\sqrt{l_2(l_1+l_2)(l_2+2l_3)(l_1+l_2+2l_3)}}{2l_2(l_2+l_2)l_3}
\end{aligned}
\right.
\label{eq17}
\end{equation}
which represent the totality of all possible cases of the system. In this simulation, $l_1=0.15m$, $l_2=0.1m$, and $l_3=0.7m$. The corresponding integrated gradients of the quadrupoles are $K_1=11.055 m^{-1}$, $K_2=-6.393 m^{-1}$, and $K_1=-11.055 m^{-1}$,  $K_2=7.821 m^{-1}$.

\begin{center}
\includegraphics[width=8.5cm]{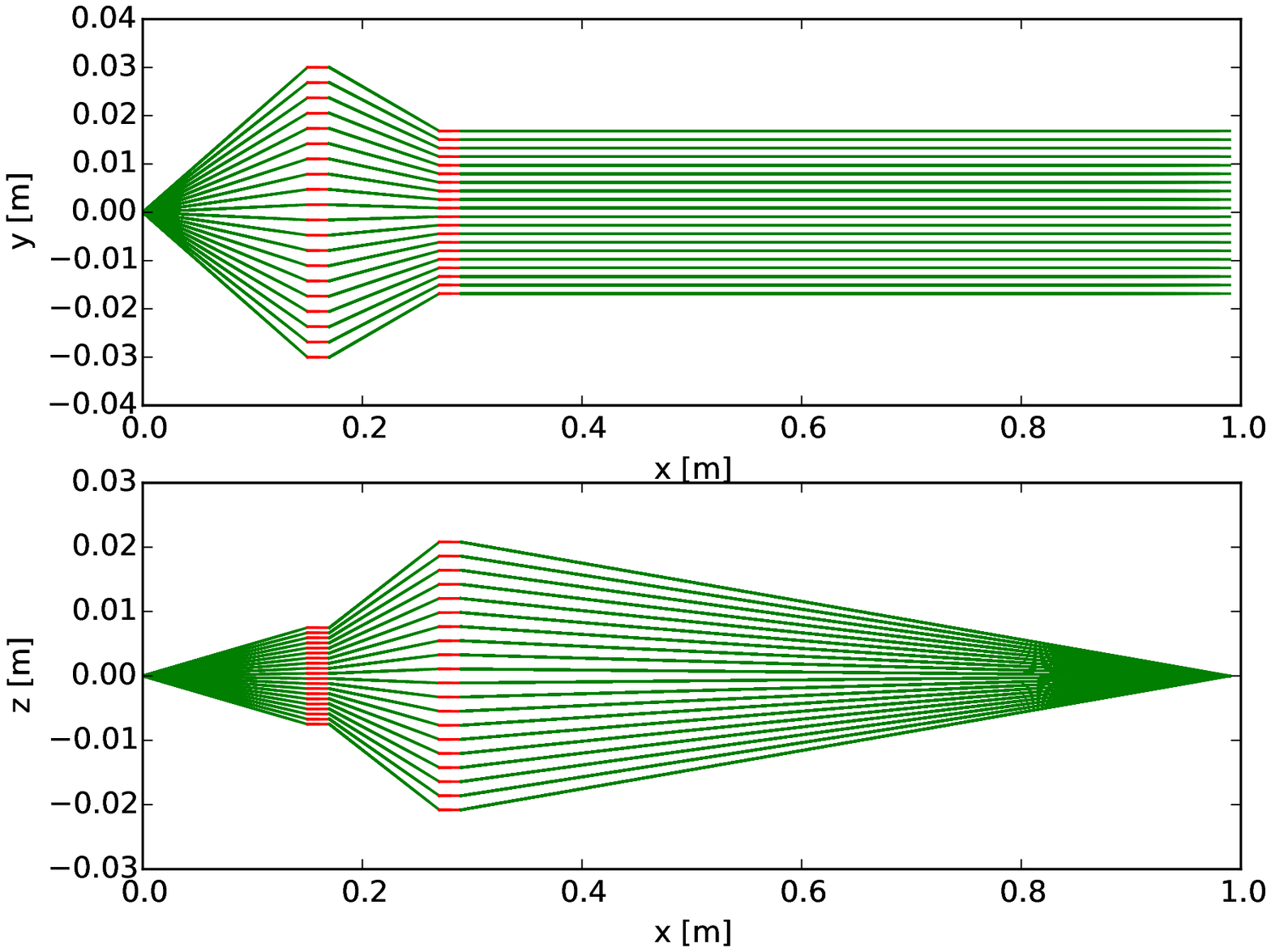}
\figcaption{\label{fig4}Top: electron trajectory in the $xy$ plane with the second solution. Bottom: electron trajectory in the $xz$ plane with the second solution.}
\end{center}

The first solution, given by Eq.~(\ref{eq16}), indicates that the first quadrupole is the defocusing one and the second is the focusing one in the $xy$ plane, and vice versa in the $xz$ plane. This can provide a wide range of parallel beam in the scanning plane and a focused beam in the other plane, as shown in Fig.~\ref{fig3}. The other solution, given by Eq.~(\ref{eq17}), also provides a parallel beam in the $xy$ plane and a focused beam in the $xz$ plane (see Fig.~\ref{fig4}), but the scan amplitude in the region of interaction is much smaller than for the first solution. Furthermore, in the $xz$ plane, the second solution is large in the region of interaction, which should be avoided to improve measurement accuracy. Therefore, the first solution seems to be better for forming parallel electron beam. We therefore focus our attention and base our simulation on the first case.

\section{PROFILE RECONSTRUCTION WITH FAST SCAN}
From Eq. (\ref{eq9}), we know that the derivative of the probe beam deflection angle with respect to impact parameter gives the projection profile of the beam cross-section distribution on the y direction. When the scan period is much shorter than the bunch length, we can perform a fast scan, so the fine structure of the bunch shape can be seen along the bunch. However, it is difficult to design a RF deflector with such high frequency, i.e. GHz. An alternate approach, the slow scan, is considered in Section~\ref{slow}. Here, we will calculate the deflection angle with a traditional approach under various beam distributions, such as KV distribution, waterbag distribution, parabolic distribution, Gaussian distribution and halo distribution. The maximum deflection occurs at the boundary for a clear-boundary beam and at the $1.585\sigma$ point for a Gaussian beam. This is verified by simulation.

The momentum change in the target beam space charge field in the $y$ direction is given by
\begin{equation}
\Delta p_y=F_y\Delta t=-e\int_{x_i}^{x_f}\frac{E_y}{v_e}dx
\label{eq18}
\end{equation}
where $x_i$ and $x_f$ represent the initial and final positions of the electron beam respectively. $E_y$ is the $y$ component of the target beam electric field and $v_e$ the velocity of the electron. Hence, the deflection angle is
\begin{equation}
\theta_y \approx \frac{\Delta p_y}{p_x}
\label{eq19}
\end{equation}
where $p_x$ is the momentum of the electron beam. If we calculate out the $y$ component of the target beam electric field, the deflection angle can be easily solved, and also the derivative. For simulation, we select a proton beam as the target beam, with kinetic energy 5~MeV/u, and number of protons per unit length, $\lambda$, of $1.87\times 10^{10}$~$m^{-1}$. The beam radius, $R$,  for the KV, waterbag and parabolic distributions is 5~mm. Considering the low current of the target beam, the electron gun cathode voltage, $V$, is selected to be 5~kV.
\subsection{KV distribution}
The KV distribution is a well-known distribution which was discovered by I. Kapchinskij and V. Vladimirskij~\cite{bib23} in 1959. The 2D real-space particle number density is defined as
\begin{equation}
n(x,y)=\frac{\lambda}{\pi R^2}, \quad x^2+y^2\le R^2
\label{eq20}
\end{equation}
where $\lambda$ is the particle density per unit length. The 1D real-space profile is given by
\begin{equation}
n(y)=\frac{2\lambda}{\pi R}\left(1-\frac{y^2}{R^2}\right)^{0.5}
\label{eq21}
\end{equation}
and the $y$ component of electric field is given by
\begin{equation}
E_y=\frac{Ze}{2\pi\epsilon_0}\gamma\lambda
\left\{
\begin{aligned}
&\frac{y}{R^2},\quad y<R\\
&\frac{y}{x^2+y^2},\quad y\ge R
\end{aligned}
\right.
\label{eq22}
\end{equation}
where $Z$ is the target beam charge, $\epsilon_0$ is the permittivity of vacuum and $\gamma$ the relativity factor. The deflection angle due to the target beam for the $y>R$ region is
\begin{equation}
\theta_y=2K\arctan \frac{\vert x_i \vert}{y}, \quad with \quad K=\frac{-Ze}{4\pi\epsilon_0 V}\gamma\lambda
\label{eq23}
\end{equation}
When $x_i \to \infty$, the anti-tangent value becomes $\arctan \frac{\vert x_i \vert}{y}\to\frac{\pi}{2}$, so the maximum deflection angle is
\begin{equation}
\theta_{max}=\vert K\pi \vert=\frac{Ze}{4\epsilon_0 V}\gamma\lambda = \frac{Z\gamma}{4\epsilon_0 v}\frac{i_b}{V}
\label{eq24}
\end{equation}
where $v$ is the velocity of the target beam and $i_b$ is the target beam current. Applying the parameters given before, the maximum deflection angle is $\theta_{max}=\frac{1.6\times 10^{-19}\times 1.87\times 10^{10}}{4\times 8.854\times 10^{-12}\times 5000}\approx 17$~mrad. For a drift distance $L=1$~m from gun exit to screen, the deflection reached 17~mm, which requires that the diameter of the screen should be 4~cm or even larger. The deflected trajectory of the electron beam and reconstructed profile of the target beam are illustrated in Fig.~\ref{fig7}.
\begin{center}
\includegraphics[width=8.5cm]{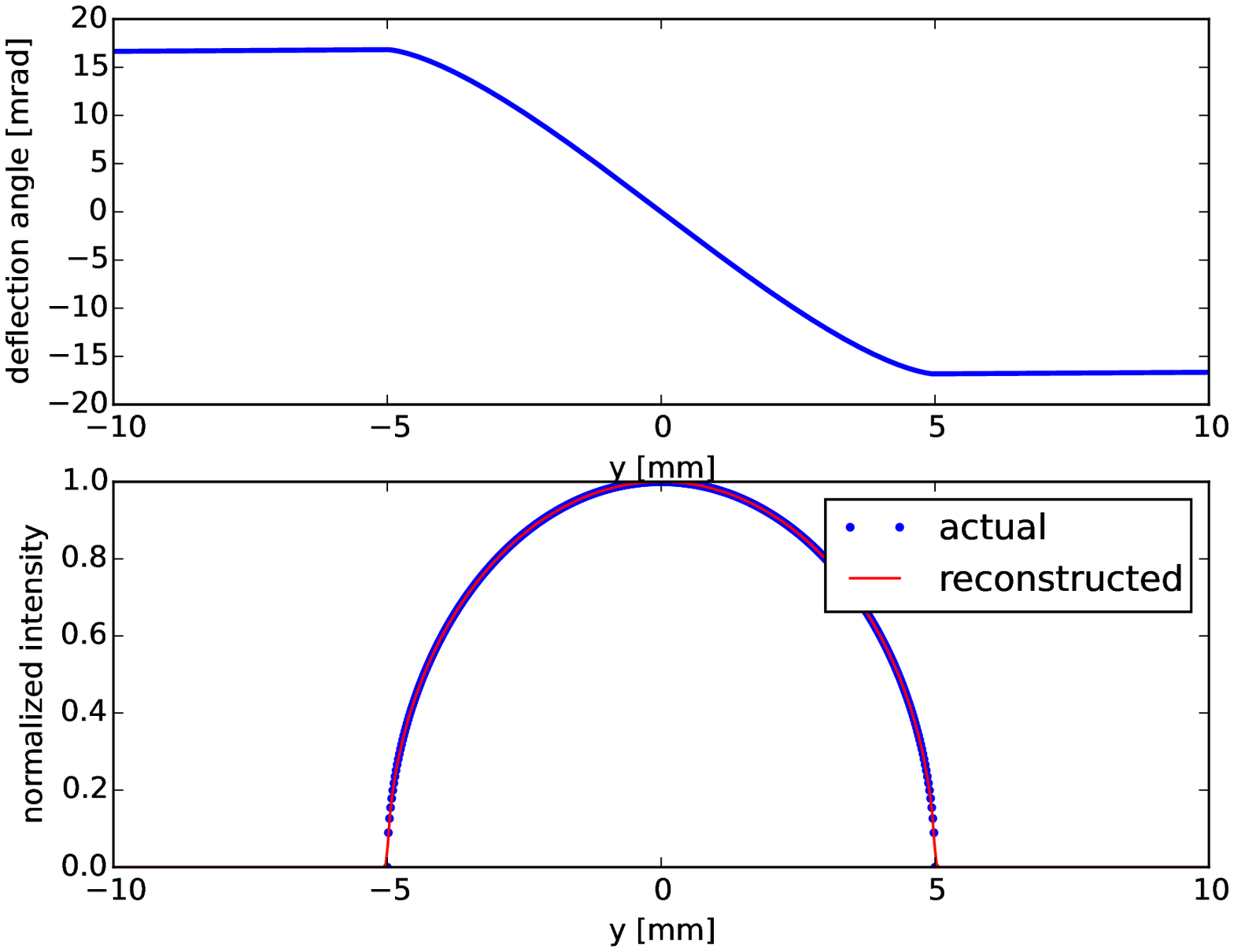}
\figcaption{\label{fig7}KV distribution. Top: deflected trajectory of electron beam. Bottom: reconstructed and actual profiles of target beam.}
\end{center}
\begin{center}
\includegraphics[width=8.5cm]{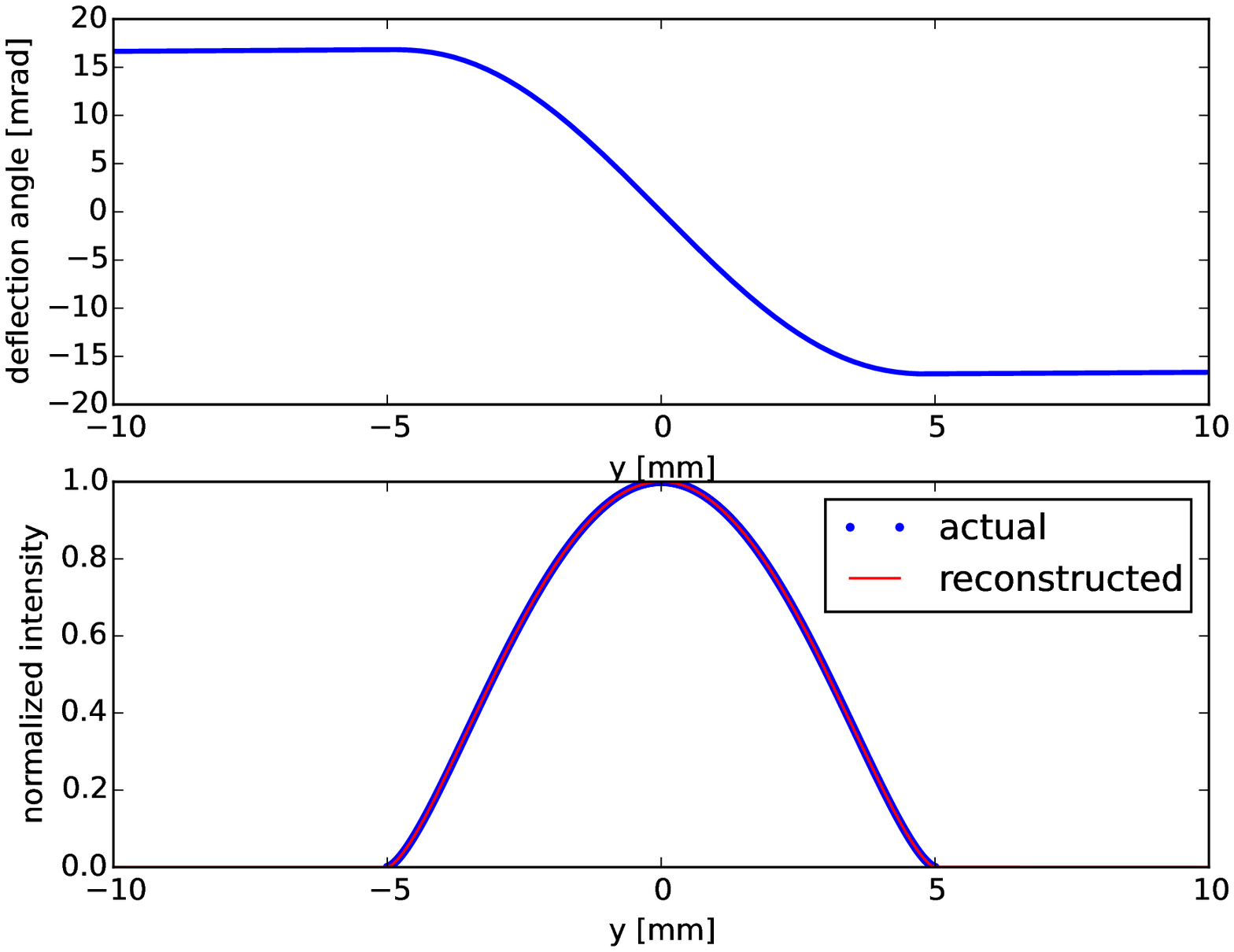}
\figcaption{\label{fig8}Waterbag distribution. Top: deflected trajectory of electron beam. Bottom: reconstructed and actual profiles of target beam.}
\end{center}
\subsection{Waterbag distribution}
The 2D real-space particle number density is defined as
\begin{equation}
n(x,y)=\frac{2\lambda}{\pi R^2}\left(1-\frac{r^2}{R^2}\right), \quad x^2+y^2\le R^2
\label{eq25}
\end{equation}
The 1D real-space profile is given by
\begin{equation}
n(y)=\frac{8\lambda}{3\pi R}\left(1-\frac{y^2}{R^2}\right)^{1.5}
\label{eq26}
\end{equation}
The $y$ component of the electric field is given by
\begin{equation}
E_y=\frac{Ze}{2\pi\epsilon_0}\gamma\lambda\frac{y}{x^2+y^2}
\left\{
\begin{aligned}
&1-\left(1-\frac{x^2+y^2}{R^2}\right)^2,\quad y<R\\
&1,\quad y\ge R
\end{aligned}
\right.
\label{eq27}
\end{equation}
The same as the KV distribution, the maximum deflection angle is also
\begin{equation}
\theta_{max}=\vert K\pi \vert=\frac{Ze}{4\epsilon_0 V}\gamma\lambda = \frac{Z\gamma}{4\epsilon_0 v}\frac{i_b}{V}
\label{eq28}
\end{equation}
which is decided by Gauss's law. The deflected trajectory of the electron beam and reconstructed profile of the target beam are illustrated in Fig.~\ref{fig8}.
\subsection{Parabolic distribution}
The 2D real-space particle number density is defined as
\begin{equation}
n(x,y)=\frac{3\lambda}{\pi R^2}\left(1-\frac{r^2}{R^2}\right)^2, \quad x^2+y^2\le R^2
\label{eq29}
\end{equation}
 and the 1D real-space profile is given by
\begin{equation}
n(y)=\frac{16\lambda}{5\pi R}\left(1-\frac{y^2}{R^2}\right)^{2.5}
\label{eq30}
\end{equation}
The $y$ component of the electric field is given by
\begin{equation}
E_y=\frac{Ze}{2\pi\epsilon_0}\gamma\lambda\frac{y}{x^2+y^2}
\left\{
\begin{aligned}
&1-\left(1-\frac{x^2+y^2}{R^2}\right)^3,\quad y<R\\
&1,\quad y\ge R
\end{aligned}
\right.
\label{eq31}
\end{equation}
The maximum deflection is identical to the kV distribution case,
\begin{equation}
\theta_{max}=\vert K\pi \vert=\frac{Ze}{4\epsilon_0 V}\gamma\lambda = \frac{Z\gamma}{4\epsilon_0 v}\frac{i_b}{V}
\label{eq32}
\end{equation}
The deflected trajectory of the electron beam and reconstructed profile of the target beam are illustrated in Fig.~\ref{fig9}.
\begin{center}
\includegraphics[width=8.5cm]{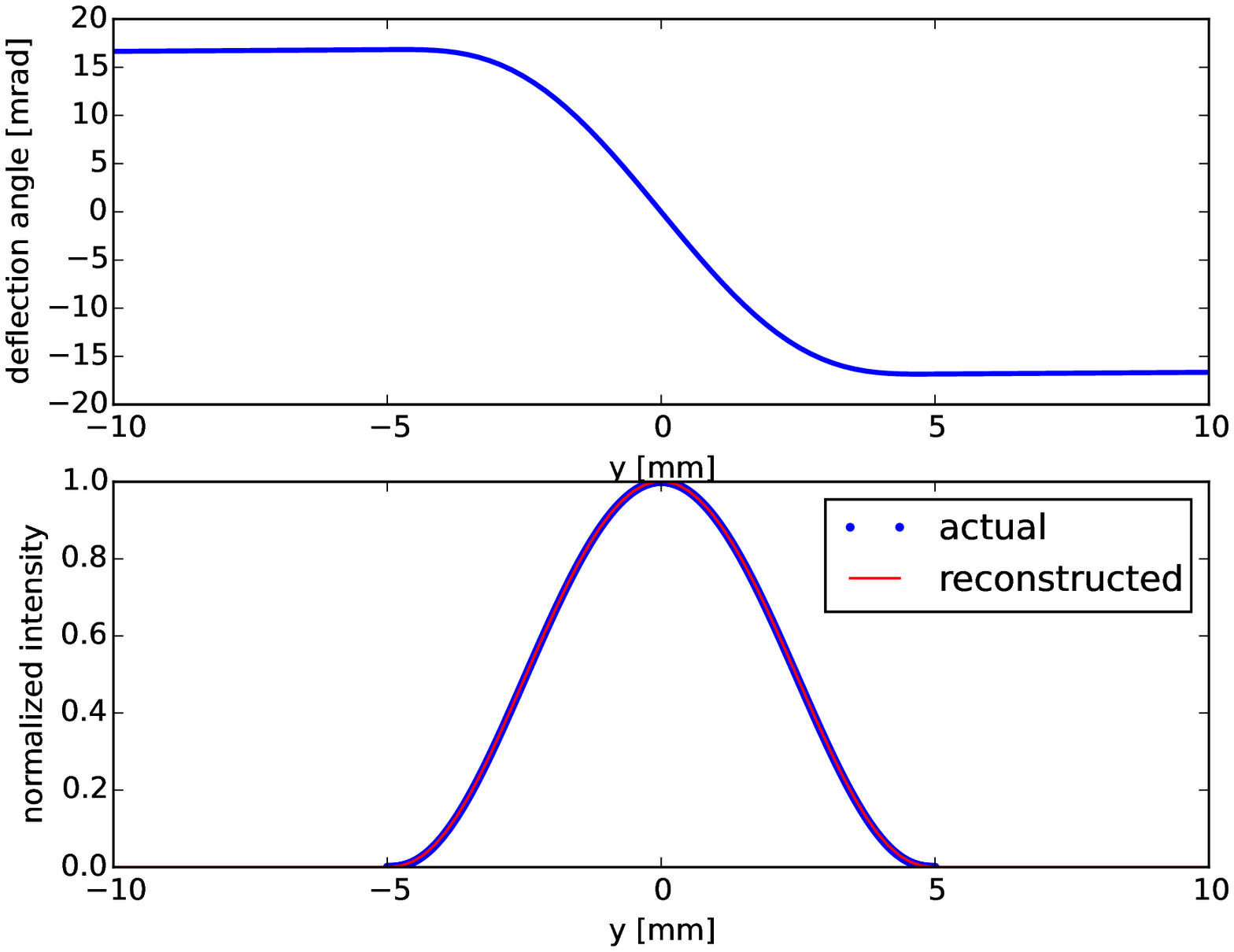}
\figcaption{\label{fig9}Parabolic distribution. Top: deflected trajectory of electron beam. Bottom: reconstructed and actual profiles of target beam.}
\end{center}
\begin{center}
\includegraphics[width=8.5cm]{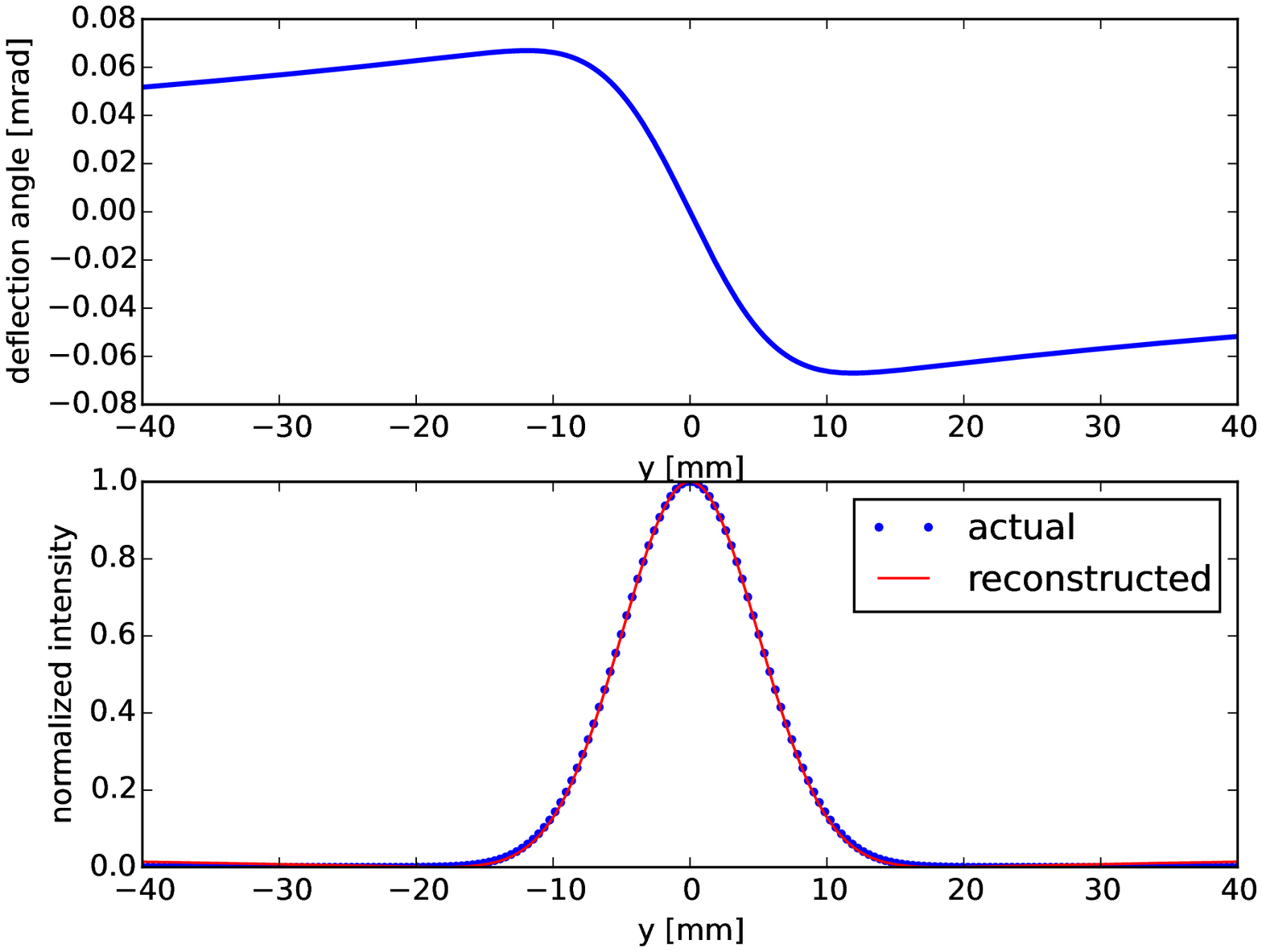}
\figcaption{\label{fig10}Gauss distribution. Top: deflected trajectory of electron beam. Bottom: reconstructed and actual profiles of target beam.}
\end{center}
\subsection{Gaussian distribution}
In this section, we present the simulation for a Gaussian distribution, which has a complex mathematical expression. The calculation of electric field can be found in Appendix A.

The 2D real-space particle number density can be formulated as
\begin{equation}
n(x,y)=\frac{\lambda}{2\pi\sigma_x\sigma_y}e^{-\frac{x^2}{2\sigma_x^2}-\frac{y^2}{2\sigma_y^2}}
\label{eq33}
\end{equation}
The profile in the $y$ direction is given by
\begin{equation}
n(y)=\frac{\lambda}{\sqrt{2\pi}\sigma_y}e^{-\frac{y^2}{2\sigma_y^2}}
\label{eq34}
\end{equation}
The $y$ component of the electric field is given by
\begin{equation}
E_y=\Phi_0\frac{2}{\pi}\frac{\gamma y}{\sigma_0}\int_\kappa^1 d\xi\frac{1}{\xi^2}\frac{1}{\sqrt{q_z}}e^{-\frac{x^2}{q_x}-\frac{y^2}{q_y}-\frac{z^2}{\gamma^2q_z}}
\label{eq35}
\end{equation}
with $\sigma_0=\sqrt{2(\sigma_x^2-\sigma_y^2)}$, $\Phi_0=\frac{Ze\lambda}{2\sqrt{\pi}\epsilon_0\sigma_0}$, $\kappa=\frac{\sigma_y}{\sigma_x}$, $q_x=q+2\sigma_x^2$, $q_y=q+2\sigma_y^2$, $q_z=q+2\sigma_z^2$, $\xi=\frac{q_y}{q_x}$, and $\gamma$ the relativistic factor. For simulation, we select $\sigma_x=7$~mm, $\sigma_y=5$~mm and $\sigma_z=10$~cm.
The deflected trajectory of the electron beam and reconstructed profile of the target beam are illustrated in Fig.~\ref{fig10}.

\subsection{Halo distribution}
The calculation of electric field is given in Appendix A.

The 2D real-space particle number density can be formulated as
\begin{equation}
n(x,y)=\frac{\lambda}{\pi ab}\left(\frac{x^2}{a^2}+\frac{y^2}{b^2}\right)e^{-\frac{x^2}{a^2}-\frac{y^2}{b^2}}
\label{eq36}
\end{equation}
The profile in the $y$ direction is given by
\begin{equation}
n(y)=\frac{2\lambda}{\sqrt{\pi}b}\frac{y^2}{b^2}e^{-\frac{y^2}{b^2}}
\label{eq37}
\end{equation}
The $y$ component of the electric field is given by
\begin{equation}
\begin{aligned}
E_y=&\frac{-Ze}{6\pi^2\epsilon_0}\gamma\lambda\int_0^\infty dq\frac{e^{-\frac{x^2}{q+a^2}-\frac{y^2}{q+b^2}-\frac{z^2}{\gamma^2(q+c^2)}}}{\sqrt{q+a^2}+\sqrt{q+b^2}+\sqrt{q+c^2}}\\
&\frac{\left(qb^2-q^2+2b^4\right)y-2b^2y^3}{\left(q+b^2\right)^3}
\end{aligned}
\label{eq38}
\end{equation}
In this simulation, the parameters are $a=1$~mm, $b=0.1$~mm, $c=5$~mm. The deflected trajectory of the electron beam and reconstructed profile of the target beam are illustrated in Fig.~\ref{fig11}.

Up to now, accurate simulation results using fast scans have been carried out with several well-known beam distributions. Based on numerous examples, we have further demonstrated the power of this method.  Next, we will explore some other interesting characteristics of EBPs.
\begin{center}
\includegraphics[width=8.5cm]{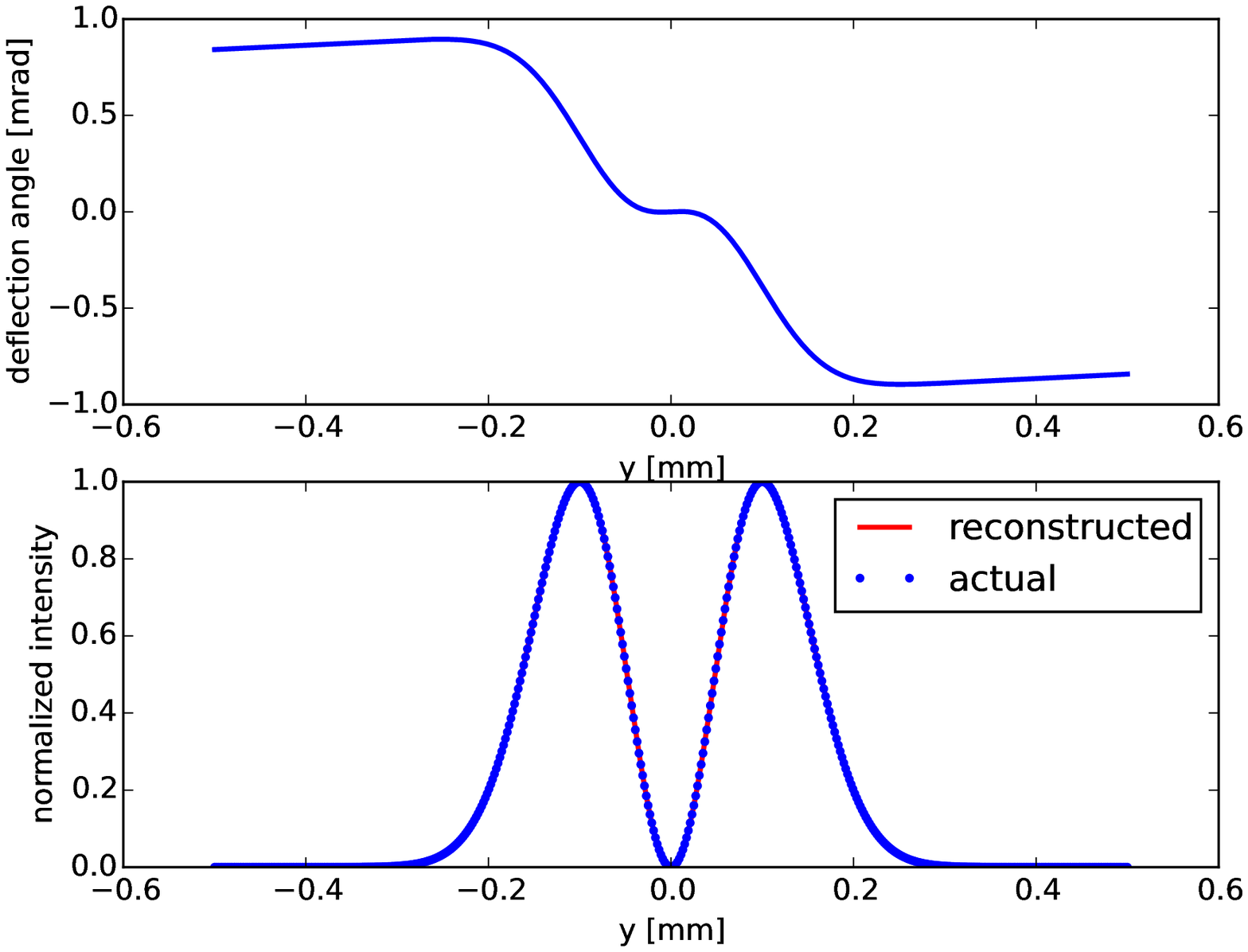}
\figcaption{\label{fig11}Halo distribution. Top: deflected trajectory of electron beam. Bottom: reconstructed and actual profiles of target beam.}
\end{center}

\section{\label{slow}PROFILE RECONSTRUCTION WITH STEP-BY-STEP SCAN}
In the former section, we investigated the fast scan and profile reconstruction with various target beam distributions. In this section, taking the Gaussian distribution as an example, we present an alternate technique to obtain the deflection angle, which is more easily  realized than the fast scan. This method involves the electron beam being slowly stepped through the target beam, while the maximum deflection angle is recorded. More specifically, the electron beam stays stationary each time the target bunch passes and is then moved to the next impact parameter by the deflector.
\begin{center}
\includegraphics[width=8.5cm]{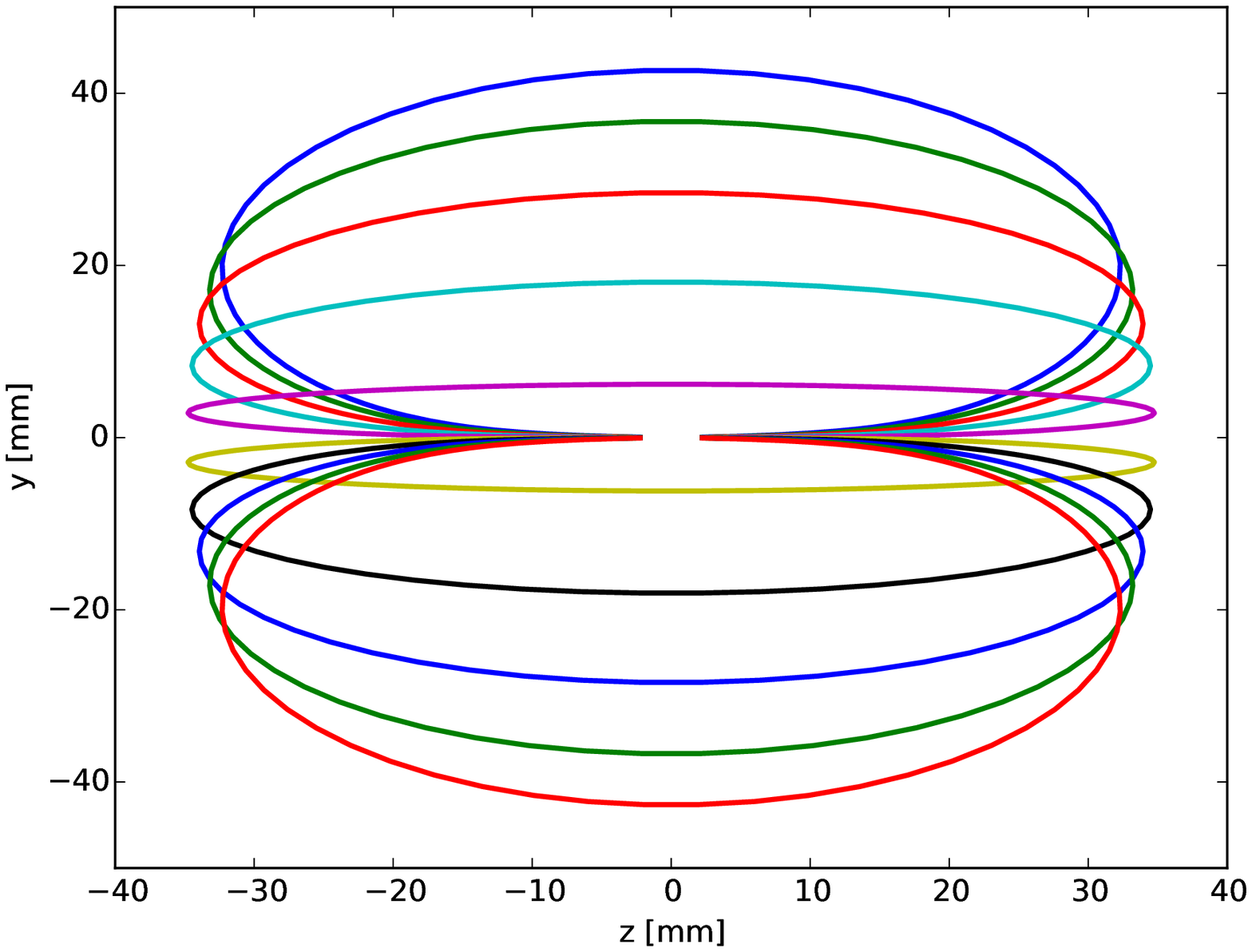}
\figcaption{\label{fig12}Step-by-step scan with varying impact parameters. From top to bottom, the impact parameters are 1~cm, 0.77~cm, 0.55~cm, 0.33~cm, 0.11~cm, -0.11~cm, -0.33~cm, -0.55~cm, -0.77~cm, -1~cm, respectively.}
\end{center}
\begin{center}
\includegraphics[width=8.5cm]{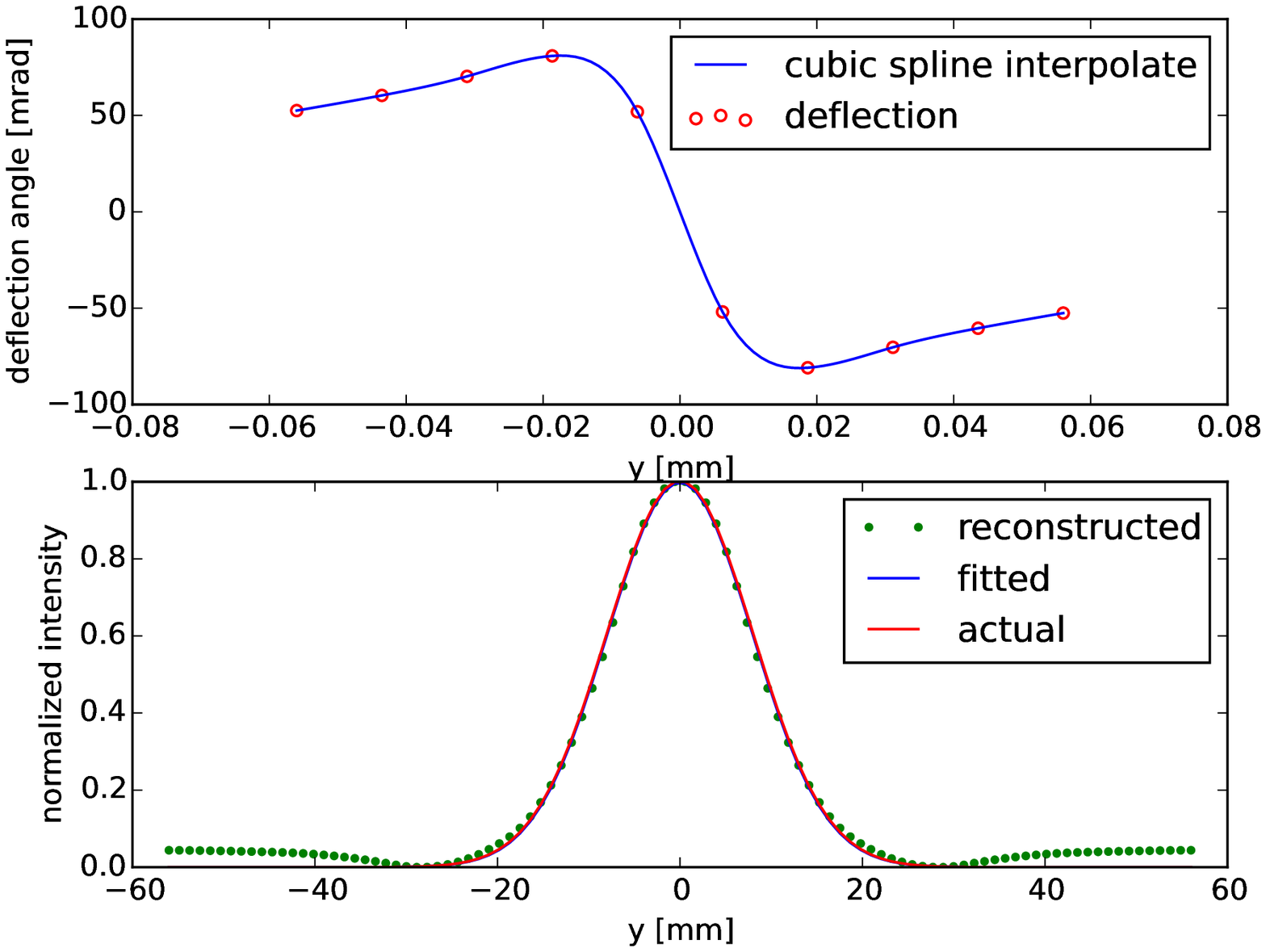}
\figcaption{\label{fig13}Deflection curve and profile reconstruction using step-by-step scan. In this case, the electron beam stays stationary when the target beam passes. Top: deflected trajectory of electron beam corresponding to each impact parameter. Bottom: reconstructed, fitted and actual profiles of target beam.}
\end{center}
\begin{center}
\includegraphics[width=8.5cm]{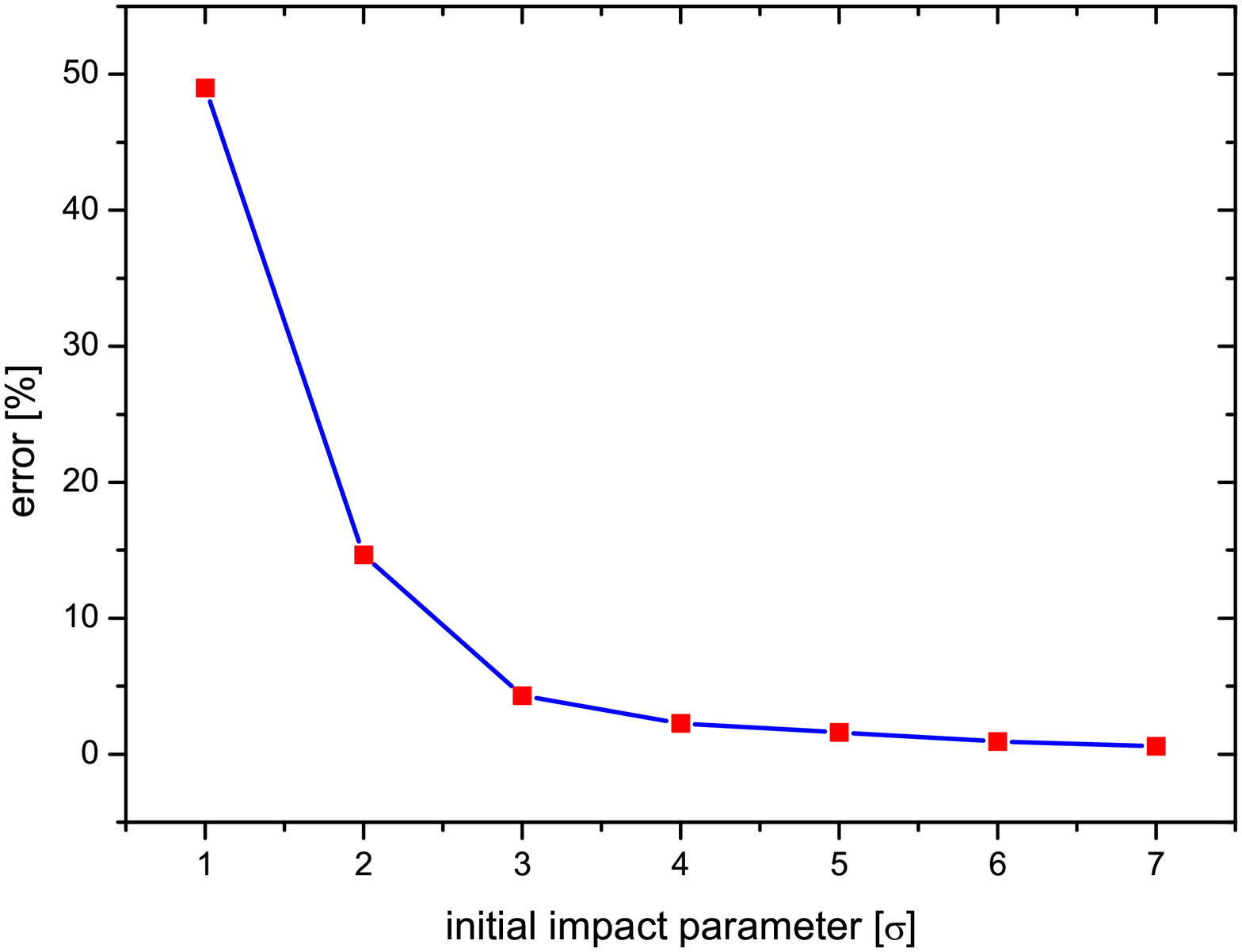}
\figcaption{\label{fig14}Profile reconstruction errors with different initial impact parameters.}
\end{center}
\begin{center}
\includegraphics[width=8.5cm]{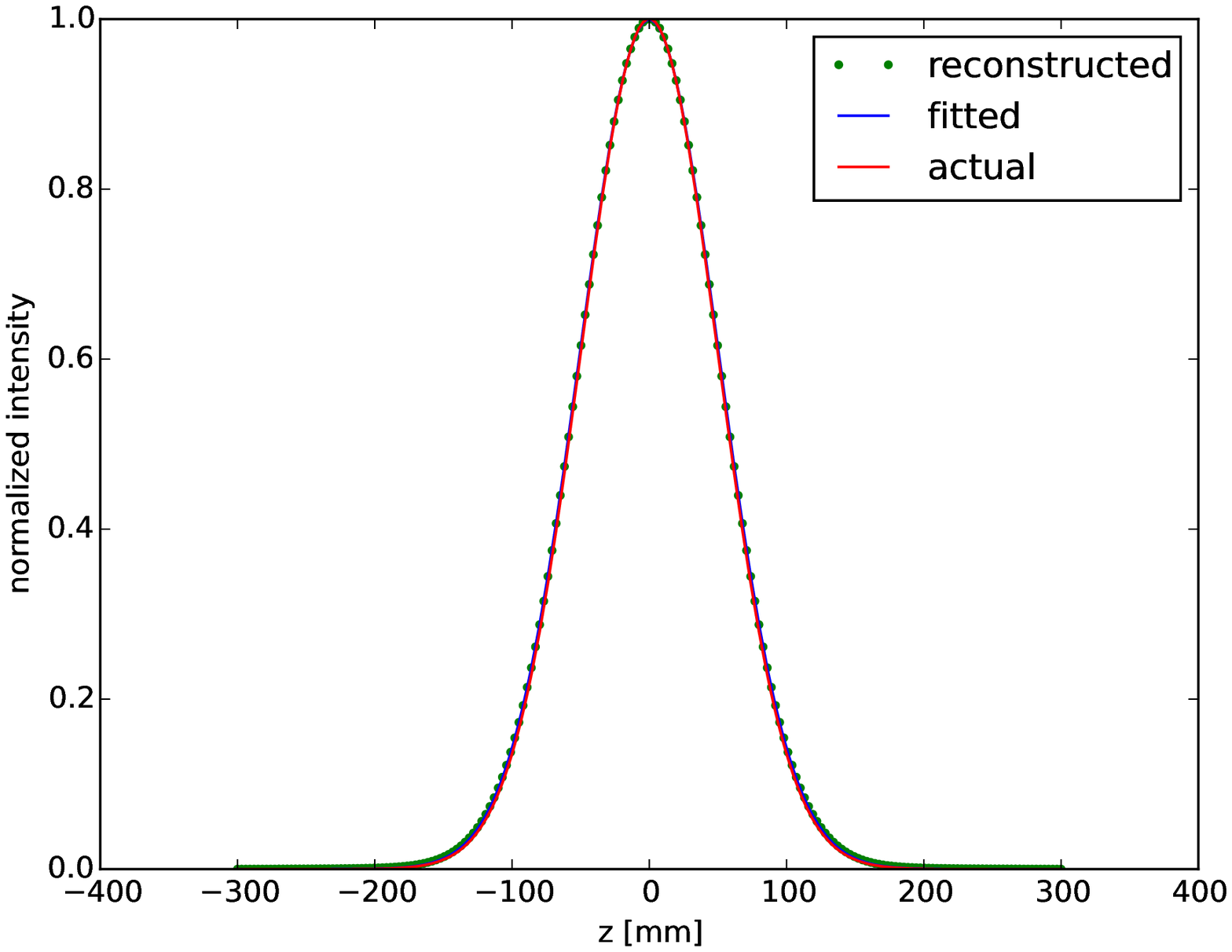}
\figcaption{\label{fig15}Reconstructed, fitted and actual bunch shapes, showing good agreement.}
\end{center}
\begin{center}
\includegraphics[width=8.5cm]{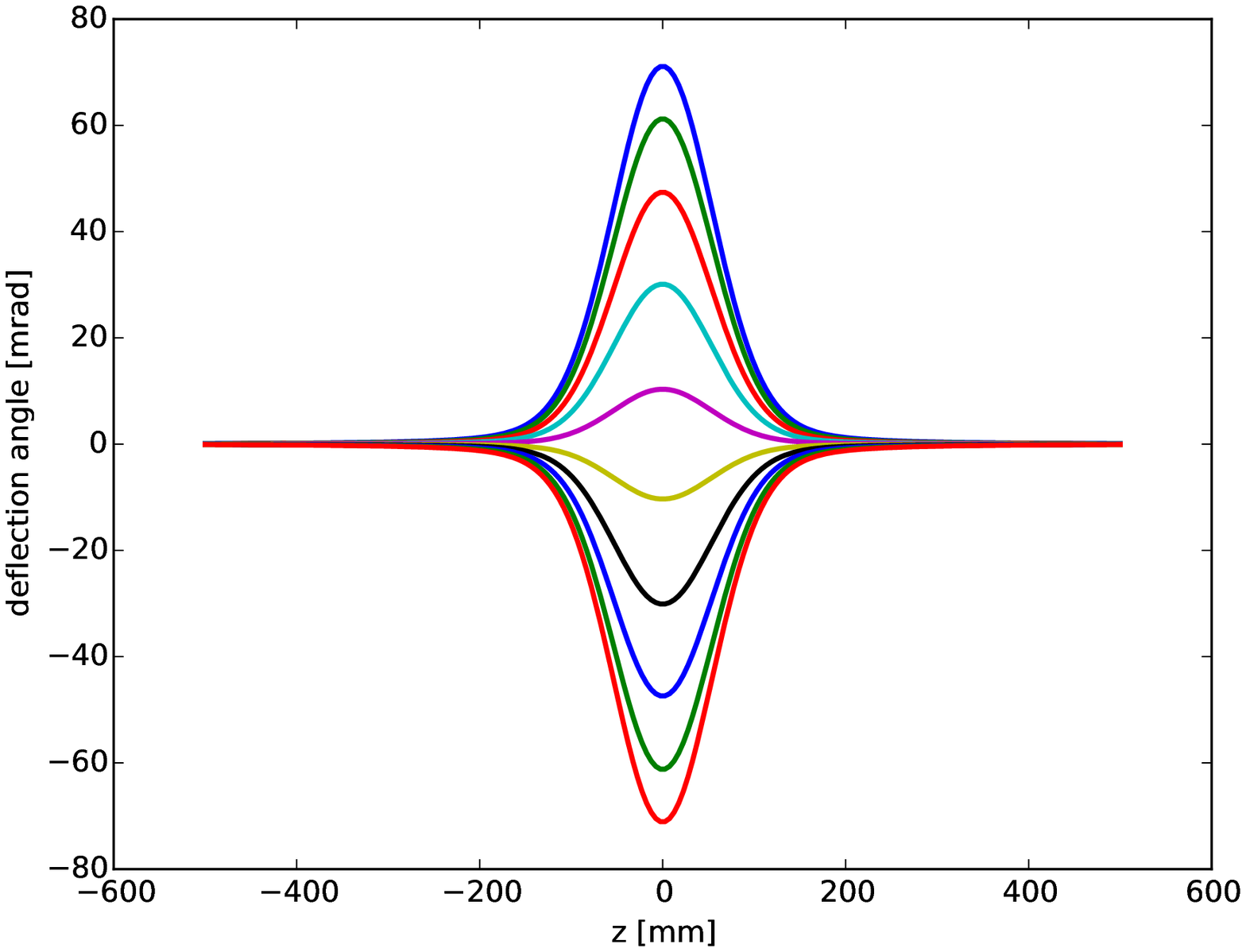}
\figcaption{\label{fig16}Scan along bunch path with varying impact parameters. From top to bottom, the impact parameters are 8~mm, 6.22~mm, 4.44~mm, 2.66~mm, 0.88~mm, -0.88~mm, -2.66~mm, -4.44~mm, -6.22~mm, -8~mm, respectively.}
\end{center}
\begin{center}
\includegraphics[width=8.5cm]{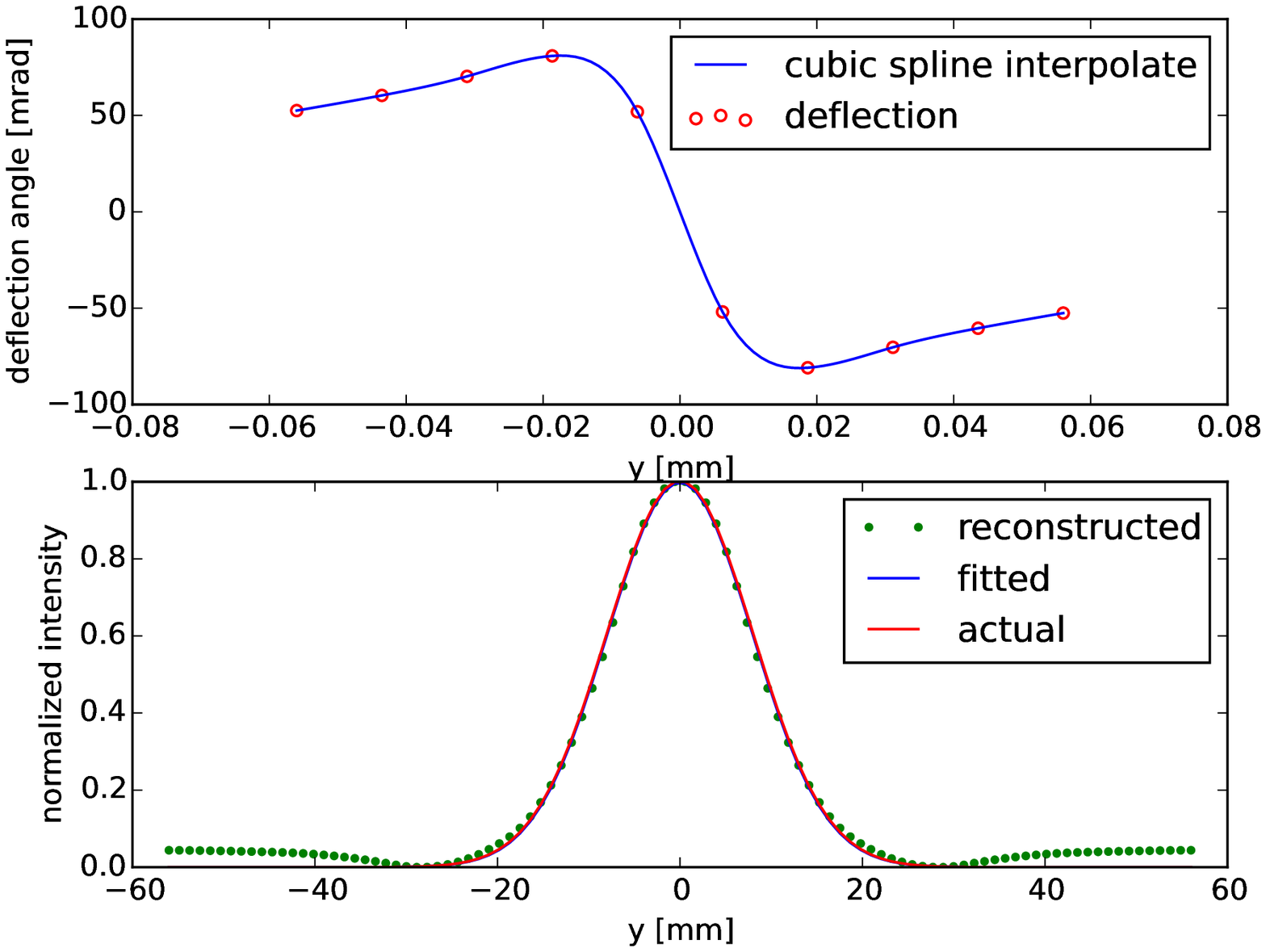}
\figcaption{\label{fig17}Deflection curve and profile reconstruction with the electron beam scanning quickly along the direction of bunch motion. Top: deflected trajectory of electron beam along bunch path with various impact parameters. Bottom: reconstructed, fitted and actual profiles of target beam.}
\end{center}

The electric field of a 3D Gaussian distribution can be decomposed into three components, $E_x$, $E_y$ and $E_z$. In some sense, $E_x$ can be neglected due to its limited role. Hence, electrons are deflected by $E_y$ and $E_z$, which can be treated in the same way as the $y$ direction was. Referring to Eq.~(\ref{eq18}), the momentum change of electrons in the $z$ direction can also be expressed as
\begin{equation}
\Delta p_z=-e\int_{x_i}^{x_f}\frac{E_z}{v_e}dx
\label{eq39}
\end{equation}
To simulate the whole procedure of electron beam motion with a 3D Gaussian bunch, we developed a Python code, which takes the transverse and longitudinal effects into account at the same time. The simulation results are illustrated in Fig.~\ref{fig12} and Fig.~\ref{fig13}.

Figure~\ref{fig12} shows the simulated deflection angle with varing impact parameters, demonstrating a symmetry about the $z$ axis, since the bunch we chose is symmetric. Figure~\ref{fig13} shows the reconstructed profile and actual profile, which agree well. In the process of simulation, we found that it is essential to scan with a large initial impact parameter, i.e. $7\sigma_y$, to obtain a nice result. The reconstruction errors corresponding to various initial impact parameters are illustrated in Fig.~\ref{fig14}.

\section{BUNCH SHAPE RECONSTRUCTION}
The fast scan along the target bunch allows measurement of the bunch structure. Furthermore, if we step the electron beam across the target bunch with varying impact parameters and record the maximum deflection angle corresponding to every deflection, the beam profile can also be extracted. To obtain a nice reconstruction, the same as for the step-by-step scan, the initial impact parameters should be large enough. Figure~\ref{fig15} shows the bunch shape achieved from the fast scan along the target bunch. In order to accurately describe and predict the bunch shape, $y$ should be as small as possible, i.e. $0.5\sigma_y$. Figure~\ref{fig18} shows the reconstruction errors corresponding to various $y$ values. The stepping scan with varying impact parameters and the reconstructed profile are illustrated in Fig.~\ref{fig16} and Fig.~\ref{fig17}, respectively.
\begin{center}
\includegraphics[width=8.5cm]{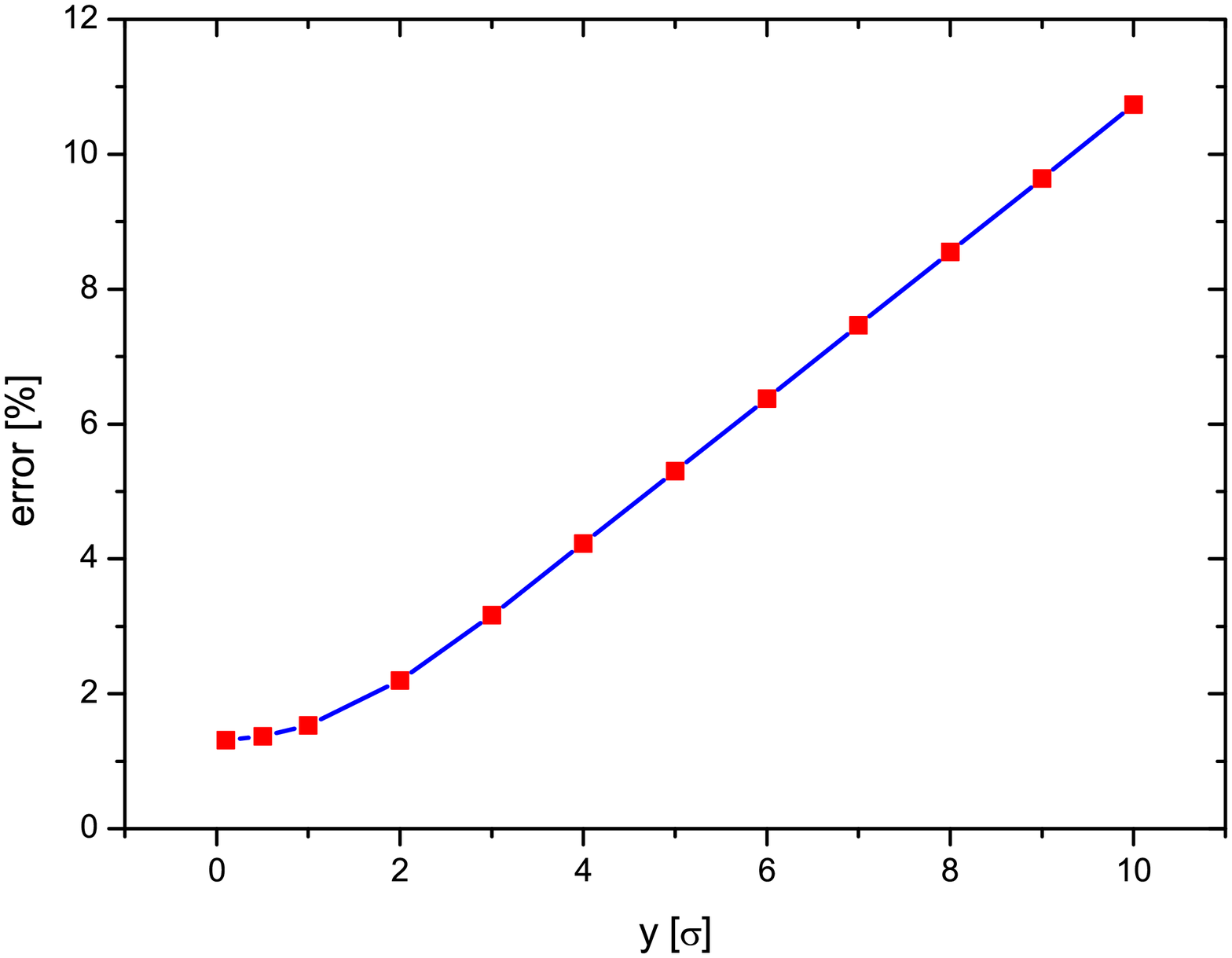}
\figcaption{\label{fig18}Bunch shape reconstruction errors with different initial values of $y$.}
\end{center}
\begin{center}
\includegraphics[width=8.5cm]{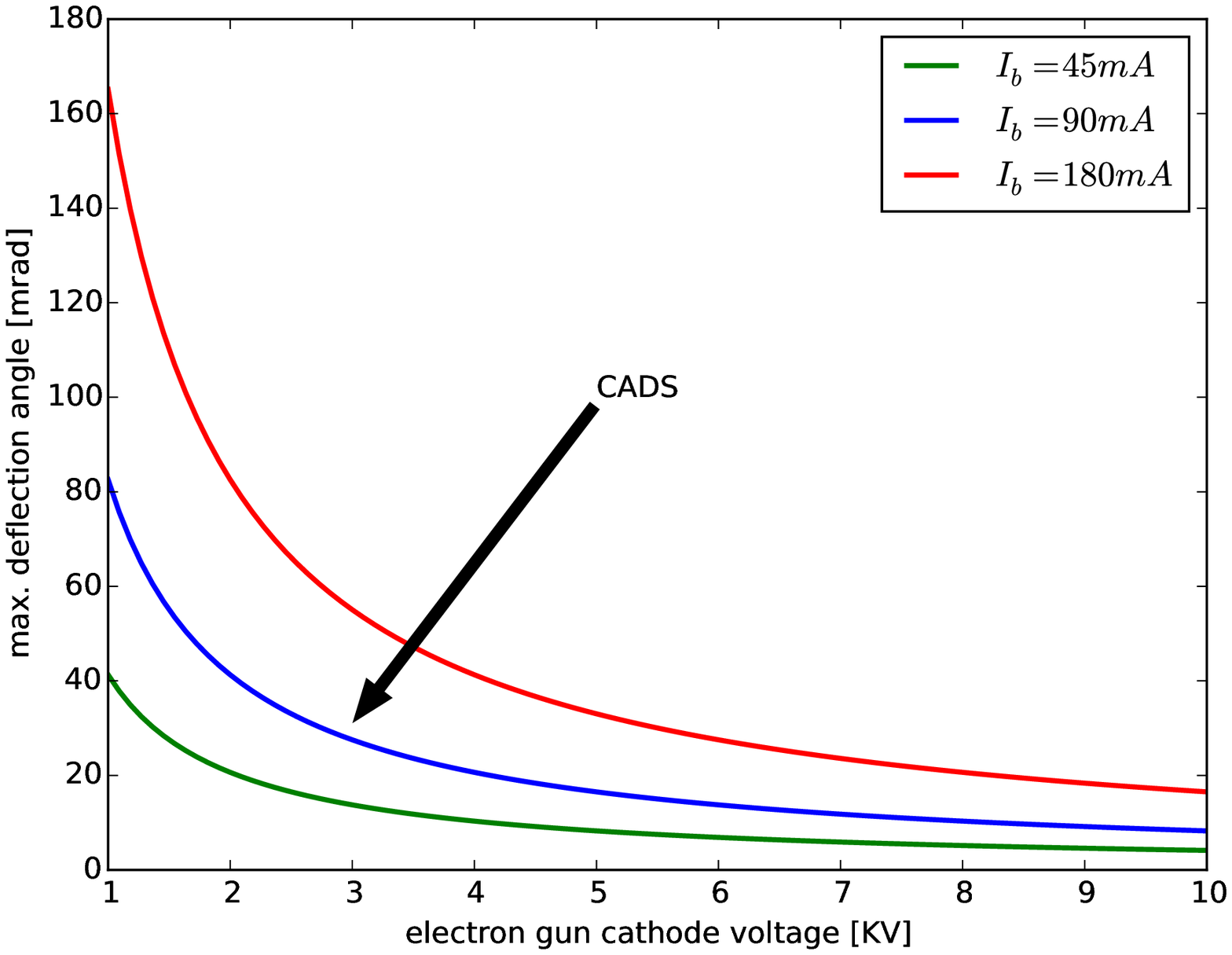}
\figcaption{\label{fig19}Maximum deflection angle with different target bunch currents under various electron gun energies.}
\end{center}
\begin{center}
\includegraphics[width=8.5cm]{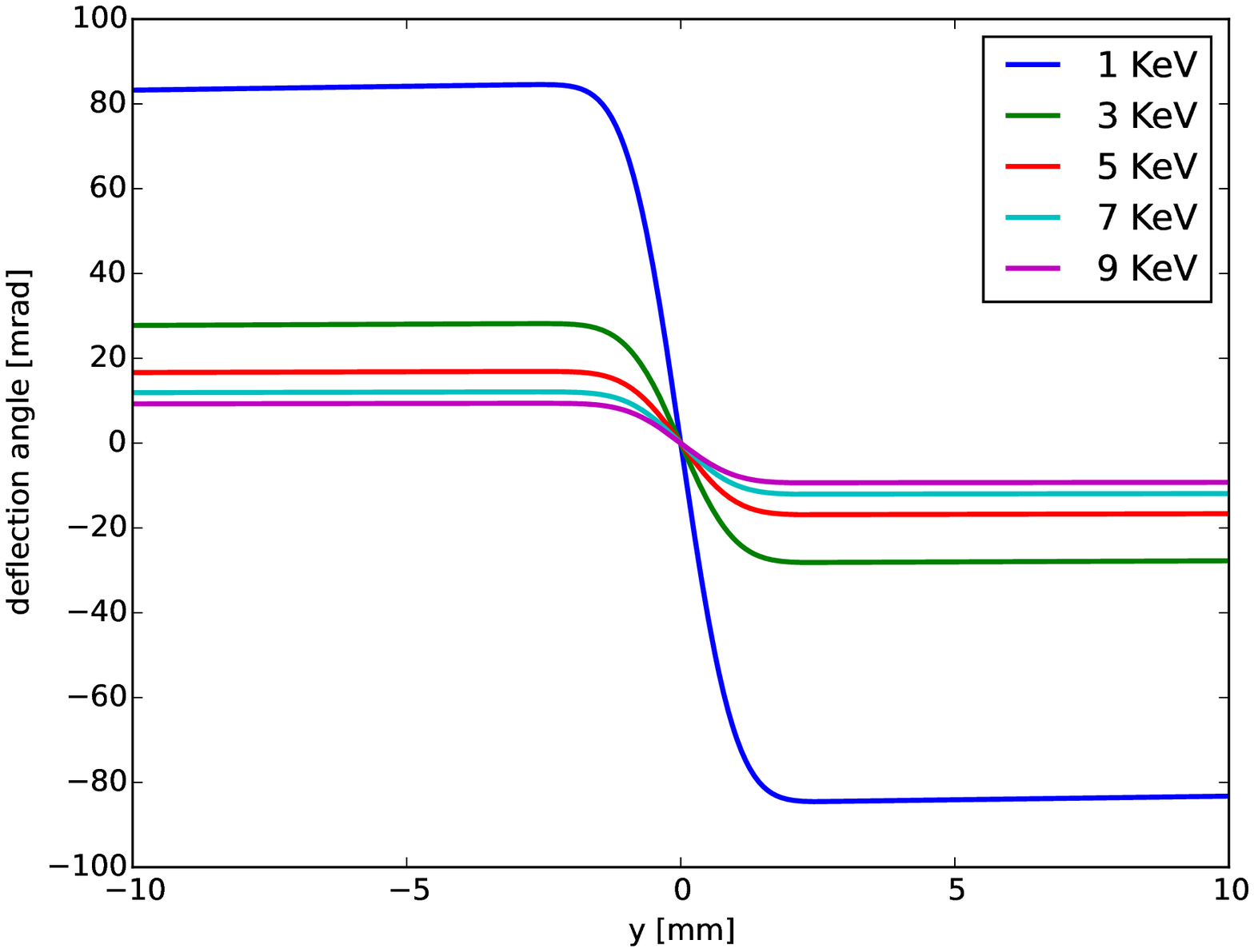}
\figcaption{\label{fig20}Deflection angle under different electron gun energies with a Gaussian target beam. Clearly, with the increase of electron gun energy, the deflection angle becomes smaller.}
\end{center}
\begin{center}
\includegraphics[width=8.5cm]{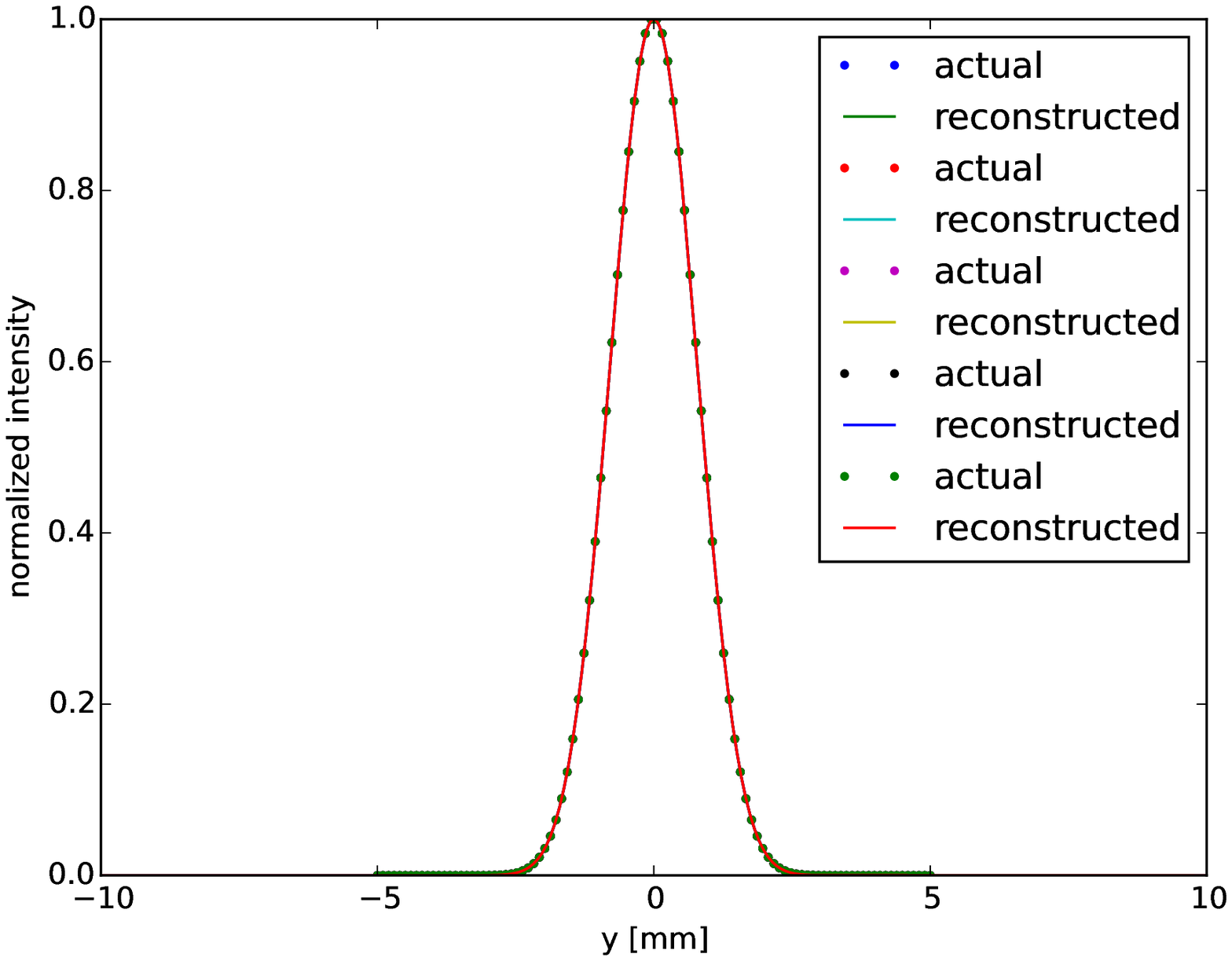}
\figcaption{\label{fig21}Reconstructed profiles from Fig.~\ref{fig20}. Although the deflection curves are extremely different for the five electron gun energies, there is good agreement between the reconstructed beam profiles and the actual profiles.}
\end{center}

\section{SIMULATION FOR CADS AND HIAF}
\paragraph{CADS}
In this section, we present an example to verify the sensitivity of the deflection angle corresponding to target beam current and the profile reconstruction under different electron gun energies. We first take the LEBT of the China Accelerator Driven Sub-critical System (CADS) as an example. It accelerates proton beam bunches to 5~MeV/u at macro pulse current of 15~mA with a macro pulse length of 500~$\mu$s. The RF frequency is 162.5~MHz and each bucket has a phase width of $\frac{1}{3}\pi$ with a micro bunch current $i_b\approx 90$~mA. Let the electron gun energy be 5~keV, so the maximum deflection angle (see Eq.~\ref{eq24}) is about 17~mrad. If the distance from bunch center to screen is 0.6~m, the diameter of the screen should be at least 21~mm. Figure~\ref{fig19} shows the maximum deflection angle with three different bunch currents, 45~mA, 90~mA and 180~mA. Considering the screen size, for the CADS LEBT current, a 5~keV electron gun is enough. Figures~\ref{fig20} and \ref{fig21} show various deflection curves with several electron energies. The reconstructed profile is exactly identical although the deflection curves are extremely different.
\paragraph{HIAF}
Merging is an advanced technology to produce supercritical atoms that is of great importance to investigate the spontaneous occurrence of electron-positron pairs in a strong Coulomb field. The HIAF project employs two synchrotrons, SRing-A and SRing-B, to form a special collider. A small angle collision is implemented to merge two coasting beams together at the interaction point. To check the merging result, the beam distribution at the moment of merging is essential. An EBP provides an effective way to reconstruct the beam distribution with arbitrary cross-section. Suppose that the two beams both have a Gaussian distribution  and an offset from the center. The 2D real-space particle number density can therefore be formulated as
\begin{equation}
n(x,y)=\frac{\lambda}{2\pi\sigma_x}e^{\frac{-x^2}{2\sigma_x^2}}\left(\frac{1-\alpha}{\sigma_{y1}}e^{\frac{-y^2}{2\sigma_{y1}^2}}+\frac{\alpha}{\sigma_{y2}}e^{\frac{-(y-y_0)^2}{2\sigma_{y2}^2}}\right)
\label{eq40}
\end{equation}
where $\alpha$ is the scale factor and $y_0$ is the offset in the y direction.
The profile in the $y$ direction is given by
\begin{equation}
n(y)=\frac{\lambda}{\sqrt{2\pi}}\left(\frac{1-\alpha}{\sigma_{y1}}e^{\frac{-y^2}{2\sigma_{y1}^2}}+\frac{\alpha}{\sigma_{y2}}e^{\frac{-(y-y_0)^2}{2\sigma_{y2}^2}}\right)
\label{eq41}
\end{equation}
The $y$ component of the electric field is given by
\begin{equation}
\begin{aligned}
E_y=&\frac{Ze}{\pi^{3/2}\epsilon_0}\gamma\lambda \left[ \frac{(1-\alpha)y}{\sigma_{01}^2}\int_{\kappa_1}^1 d\xi_1\frac{q_x}{q_{y1}}e^{-\frac{x^2}{q_x}-\frac{y^2}{q_{y1}}}\right. \\
&\left. +\frac{\alpha (y-y_0)}{\sigma_{02}^2}\int_{\kappa_2}^1 d\xi_2\frac{q_x}{q_{y2}}e^{-\frac{x^2}{q_x}-\frac{(y-y_0)^2}{q_{y2}}} \right]
\end{aligned}
\label{eq42}
\end{equation}
with $q_x=\frac{\sigma_{01}^2}{1-\xi_1^2}$, $\sigma_{01}=\sqrt{2(\sigma_x^2-\sigma_{y1}^2)}$, $\kappa_1=\frac{\sigma_{y1}}{\sigma_x}$, $q_{y1}=\frac{\sigma_{01}^2\xi_1^2}{1-\xi_1^2}$,
$\sigma_{02}=\sqrt{2(\sigma_x^2-\sigma_{y2}^2)}$, $\kappa_2=\frac{\sigma_{y2}}{\sigma_x}$, and $q_{y2}=\frac{\sigma_{02}^2\xi_2^2}{1-\xi_2^2}$. In this simulation, $\sigma_x=7$~mm, $\sigma_{y1}=5$~mm, $\sigma_{y2}=6$~mm, $y_0=2$~cm, and $\alpha=0.2$.
The deflected trajectory of the electron beam and the reconstructed profile of the target beam are illustrated in Fig.~\ref{fig22}. It shows good agreement, and also verifies that the EBP is able to detect the profile of merging beams.
\begin{center}
\includegraphics[width=8.5cm]{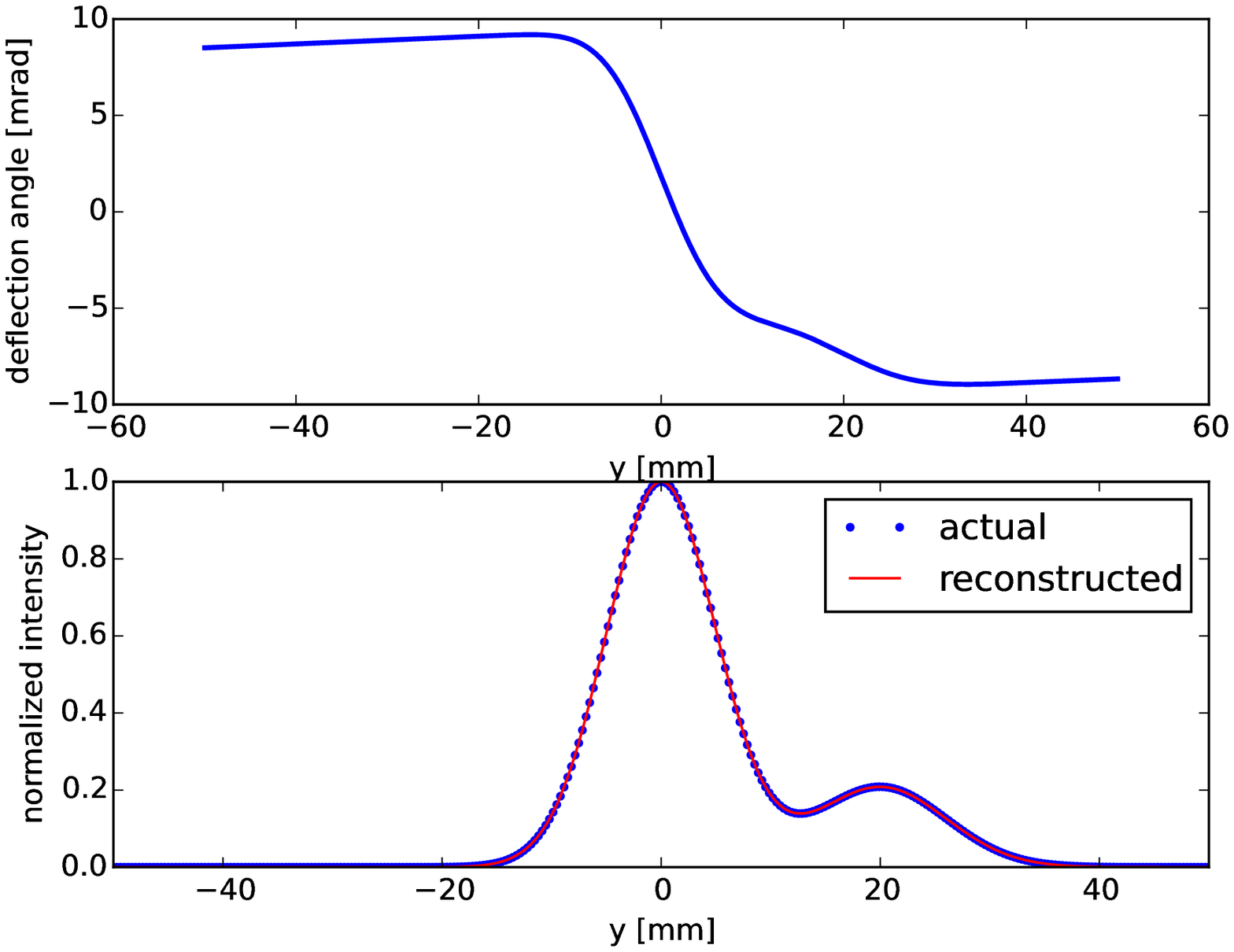}
\figcaption{\label{fig22}Bi-Gauss distribution. Top: deflected trajectory of electron beam. Bottom: reconstructed and actual profiles of target beam.}
\end{center}
\section{A POTENTIAL EBP SYSTEM DESIGN}
Based on the simulation given above, a potential system design for an electron beam probe is proposed as illustrated in Fig.~\ref{fig23}. At the electron gun exit, a solenoid is employed to make electron beam profile as small as possible in the screen. Then, a pair of deflectors, horizontal and vertical, are applied to manipulate the electron beam. To measure the vertical profile of the target beam, the vertical deflector should be powered by a RF voltage. The RF frequency depends on the frequency of the target beam if fast scan is implemented. In general, both the deflectors can be powered by a static voltage to calibrate the electron beam to pass the doublet center, e.g. by BBA (beam based alignment). The quadrupole doublet is arranged so that a parallel electron beam is produced in the scanning direction. For offline testing, a current-carrying wire can be applied to simulate the target beam. The distance from the target beam to the screen is related to electron gun energy and target beam intensity. In principle, the deflected curve should stay in the screen area, and the deflection angle should not be too small, to improve resolution. So, it is necessary to estimate these parameters in advance. We take a simple example to further understand this process. Assume the particle number per unit length of target beam is $\lambda=1.87\times 10^{10}$, and the distance from the target beam to the screen is 50~cm. To reach a maximum deflection of 1~cm on the screen, the required deflection angle is 20~mrad. Referring to the maximum deflection angle formula, Eq.~\ref{eq24}, we know that electron gun energy is less than 5~keV. Since the electron beam is easily influenced by stray magnetic fields, e.g. the geomagnetic field, especially for low energy electron beams, magnetic shielding is essential for the whole system.
\begin{center}
\includegraphics[width=8.5cm]{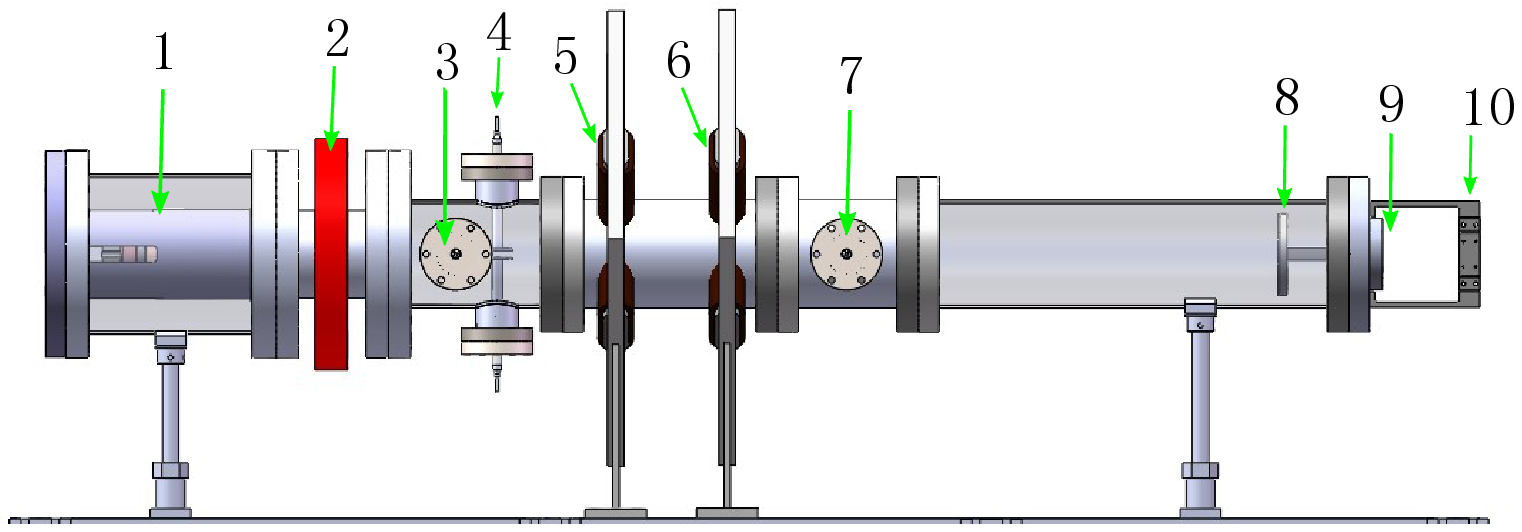}
\figcaption{\label{fig23}Layout of EBP (side view): (1) electron gun; (2) solenoid; (3) horizontal deflector; (4) vertical deflector; (5) and (6) quadrupole doublet; (7) test wire or beam; (8) screen; (9) view window; (10) CCD holder.}
\end{center}
\section{CONCLUSIONS}
In this paper, the EBP as a profile monitor and bunch length detector has been simulated comprehensively for the first time. The theoretical aspects of the technique were analyzed in detail without loss of generality. A method to produce parallel electron beams was introduced. A Python code has been developed to simulate the production of parallel electron beams with arbitrary arrangements of quadrupole doublet. Via fast scan, transverse profile reconstruction has been implemented under various well known beam distributions, such as KV distribution, waterbag distribution, parabolic distribution, Gaussian distribution and halo distribution, with good agreement. To reduce the requirement for hardware, a slow scan is proposed, which also can obtain a nice result via the presented simulation. The bunch shape can be reconstructed from the electron beam deflection along the bunch path. Meanwhile, if we vary the impact parameters continuously, the transverse profile can also be extracted. As an example, we have also shown how to select an electron gun under different target beam currents, and verified that it is possible to reconstruct the profile for merging beams. Finally, a potential system design has been presented to put the theory into practice.  Next, we plan to do more studies and fabricate a prototype EBP to verify the principle and confirm the measurement accuracy.
\section{\label{appendix}APPENDIX}
The electric field described by Eq.~(\ref{eq35}) and Eq.~(\ref{eq38}) can be computed using a method similar to that of K. Takayama~\cite{bib24,bib25,bib26}. The Poisson equation with charge distribution $\rho(\vec r)=\rho(\vec r^\prime)\delta(\vec r-\vec r^\prime)$ can be expressed as
\begin{equation}
\nabla^2\Phi(\vec r)=-\frac{\rho(\vec r)}{\epsilon_0}
\label{eqa1}
\end{equation}
The Green function corresponding to Eq.~(\ref{eqa1}) is well known in the form
\begin{equation}
G(\vec r, \vec r^\prime)=\frac{1}{4\pi\vert \vec r-\vec r^\prime\vert}
\label{eqa2}
\end{equation}
which satisfies the Green equation
\begin{equation}
\nabla^2 G(\vec r, \vec r^\prime)=-\delta(\vec r-\vec r^\prime)
\label{eqa3}
\end{equation}
and the general expression of potential can be formulated as
\begin{equation}
\Phi(\vec r)=\frac{1}{\epsilon_0}\int\int\int G(\vec r, \vec r^\prime)d^3\vec r^\prime
\label{eqa4}
\end{equation}
To apply the Green function to a Gaussian charge distribution conveniently, we can rewrite Eq.~(\ref{eqa2}) as an integral representation
\begin{equation}
G(\vec r, \vec r^\prime)=\frac{1}{4\pi^{3/2}}\int_0^\infty dq\frac{1}{q^{3/2}} e^{-\frac{\vert \vec r-\vec r^\prime\vert^2}{q}}
\label{eqa5}
\end{equation}
So, the potential expressed by the Green function is
\begin{equation}
\Phi(\vec r)=\frac{1}{4\pi\epsilon_0}\int_0^\infty dq\frac{1}{q^{3/2}\sqrt{\pi}} \int d^3\vec r^\prime\rho(\vec r^\prime)e^{-\frac{\vert \vec r-\vec r^\prime\vert^2}{q}}
\label{eqa6}
\end{equation}
In principle, the potential generated by any Gauss-like charge distribution can be solved by the formula.
\paragraph{3D Gaussian distribution}
The charge density for a 3D Gaussian distribution is
\begin{equation}
\rho(x,y,z)=\frac{Q}{\left(2\pi\right)^{3/2}\sigma_x\sigma_y\sigma_z}e^{-\left(\frac{x^2}{2\sigma_x^2}+\frac{y^2}{2\sigma_y^2}+\frac{z^2}{2\sigma_z^2}\right)}
\label{eqa7}
\end{equation}
where $Q=ZeN$ and $N$ is the total particle number.

For simplicity, we firstly calculate the $x$ direction component.
\begin{equation}
I_x=\int_{-\infty}^\infty\frac{1}{\sqrt{2\pi}\sigma_x}e^{-\frac{x^{\prime 2}}{2\sigma_x^2}}\frac{1}{\sqrt{q}}e^{-\frac{\vert \vec x-\vec x^\prime\vert^2}{q}}dx^\prime
\label{eqa8}
\end{equation}
After some simple mathematical treatment, we have
\begin{equation}
I_x=\frac{1}{\sqrt{q+2\sigma_x^2}}e^{-\frac{x^2}{q+2\sigma_x^2}}
\label{eqa9}
\end{equation}
Doing the same as we did above, the electric potential of a 3D Gaussian distribution can be expressed as
\begin{equation}
\Phi(x,y,z)=\frac{Q}{4\pi\epsilon_0}\frac{1}{\sqrt{\pi}}\int_0^\infty dq\frac{1}{q_xq_yq_z}e^{-\frac{x^2}{q_x}-\frac{y^2}{q_y}-\frac{z^2}{q_z}}
\label{eqa9}
\end{equation}
where $q_x=q+2\sigma_x^2$, $q_y=q+2\sigma_y^2$ and $q_z=q+2\sigma_z^2$.
To improve computing speed, we follow the method of the well known Bassetti-Erskine formula~\cite{bib27}, which is also used by R. Wanzenberg~\cite{bib28} to simulate nonlinear motion of a point charge in the 3D space charge field of a Gaussian bunch. A formula can often be considerably simplified by a suitable transformation of variables.

We replacing the old integration variable $q$ with the new one $\xi$ in the following way
\begin{equation}
\xi^2=\frac{q_y}{q_x}=\frac{q+2\sigma_y^2}{q+2\sigma_x^2}
\label{eqa10}
\end{equation}
or
\begin{equation}
q=\frac{2\sigma_x^2-2\sigma_y^2}{1-\xi^2}-2\sigma_x^2
\label{eqa11}
\end{equation}
since $q\in(0,\infty)$, $\xi\in(\kappa,1)$.
Assume $\sigma_x>\sigma_y$ and introduce two quantities
\begin{equation}
\kappa=\frac{\sigma_y}{\sigma_x}, \quad \sigma_0=\sqrt{2\left(\sigma_x^2-\sigma_y^2\right)}
\label{eqa12}
\end{equation}
The relation between $q$ and $\xi$ can be written as
\begin{equation}
dq=\frac{2q_x^{3/2}\sqrt{q_y}}{\sigma_0^2}d\xi
\label{eqa13}
\end{equation}
Therefore, the electric potential of a 3D Gaussian distribution in terms of these variables is
\begin{equation}
\Phi(x,y,z)=\frac{Q}{4\pi\epsilon_0}\frac{2}{\sqrt{\pi}}\frac{1}{\sigma_0^2}\int_\kappa^1 d\xi\frac{q_x}{\sqrt{q_z}}e^{-\frac{x^2}{q_x}-\frac{y^2}{q_y}-\frac{z^2}{q_z}}
\label{eqa14}
\end{equation}
with $q_x=\frac{\sigma_0^2}{1-\xi^2}$, $q_y=\frac{\sigma_0^2\xi^2}{1-\xi^2}$ and $q_z=2\left(\sigma_z^2-\sigma_x^2\right)+\frac{\sigma_0^2}{1-\xi^2}$
To compare with the Bassetti-Erskine formula, we define a constant
\begin{equation}
\Phi_0=\frac{Q}{2\sqrt{\pi}\epsilon_0\sigma_0}
\label{eqa15}
\end{equation}
The potential is now
\begin{equation}
\Phi(x,y,z)=\Phi_0\frac{1}{\pi}\frac{1}{\sigma_0}\int_\kappa^1 d\xi\frac{q_x}{\sqrt{q_z}}e^{-\frac{x^2}{q_x}-\frac{y^2}{q_y}-\frac{z^2}{q_z}}
\label{eqa16}
\end{equation}
The $y$ component of the electric field can easily be  obtained:
\begin{equation}
\begin{aligned}
E_y&=-\frac{\partial}{\partial y}\Phi(x,y,z)\\
&=\Phi_0\frac{2}{\pi}\frac{y}{\sigma_0}\int_\kappa^1 d\xi\frac{1}{\xi^2}\frac{1}{\sqrt{q_z}}e^{-\frac{x^2}{q_x}-\frac{y^2}{q_y}-\frac{z^2}{q_z}}
\end{aligned}
\label{eqa17}
\end{equation}
In the laboratory frame,
\begin{equation}
E_y=\Phi_0\frac{2}{\pi}\frac{\gamma y}{\sigma_0}\int_\kappa^1 d\xi\frac{1}{\xi^2}\frac{1}{\sqrt{q_z}}e^{-\frac{x^2}{q_x}-\frac{y^2}{q_y}-\frac{z^2}{\gamma^2 q_z}}
\label{eqa17}
\end{equation}

\paragraph{3D halo distribution}
The charge density for a 3D halo distribution~\cite{bib26} is
\begin{equation}
\rho(x,y,z)=\frac{2Q}{3\pi^{3/2}abc}\left(\frac{x^2}{a^2}+\frac{y^2}{b^2}+\frac{z^2}{c^2}\right)e^{-\frac{x^2}{a^2}-\frac{y^2}{b^2}-\frac{z^2}{c^2}}
\label{eqa18}
\end{equation}
Substituting Eq.~(\ref{eqa18}) into Eq.~(\ref{eqa6}), we obtain
\end{multicols}
\begin{equation}
\Phi(x,y,z)=\frac{1}{4\pi\epsilon_0}\frac{2Q}{3\pi^{3/2}abc}\int_0^\infty dq\frac{1}{q^{3/2}\sqrt{\pi}}\int d^3r^\prime\left(\frac{x^{\prime 2}}{a^2}+\frac{y^{\prime 2}}{b^2}+\frac{z^{\prime 2}}{c^2}\right)
e^{-\left(\frac{x^{\prime 2}}{a^2}+\frac{\left(x-x^\prime\right)^2}{q}\right)-\left(\frac{y^{\prime 2}}{b^2}+\frac{\left(y-y^\prime\right)^2}{q}\right)-\left(\frac{z^{\prime 2}}{c^2}+\frac{\left(z-z^\prime\right)^2}{q}\right)}
\label{eqa19}
\end{equation}
\begin{multicols}{2}
For simplicity, we first consider the $x$ direction:
\end{multicols}
\begin{equation}
I_x=\frac{1}{4\pi\epsilon_0}\frac{2Q}{3\pi^{3/2}abc}\int_0^\infty dq\frac{1}{q^{3/2}\sqrt{\pi}}\int\int\int \frac{x^{\prime 2}}{a^2}
e^{-\left(\frac{x^{\prime 2}}{a^2}+\frac{\left(x-x^\prime\right)^2}{q}\right)-\left(\frac{y^{\prime 2}}{b^2}+\frac{\left(y-y^\prime\right)^2}{q}\right)-\left(\frac{z^{\prime 2}}{c^2}+\frac{\left(z-z^\prime\right)^2}{q}\right)}dx^\prime dy^\prime dz^\prime
\label{eqa20}
\end{equation}
\begin{multicols}{2}
After some mathematical treatment, we have
\end{multicols}
\begin{equation}
I_x=\frac{Q}{6\pi^2\epsilon_0}\int_0^\infty dq\frac{e^{-\frac{x^2}{q+a^2}-\frac{y^2}{q+b^2}-\frac{z^2}{q+c^2}}}{\sqrt{q+a^2}+\sqrt{q+b^2}+\sqrt{q+c^2}}\left[\frac{q}{2}\frac{1}{q+a^2}+
\left(\frac{ax}{q+a^2}\right)^2\right]
\label{eqa21}
\end{equation}
\begin{multicols}{2}
Therefore, the electric potential is now
\end{multicols}
\begin{equation}
\begin{aligned}
\Phi(\vec r)=&\frac{Q}{6\pi^2\epsilon_0}\int_0^\infty dq\frac{e^{-\frac{x^2}{q+a^2}-\frac{y^2}{q+b^2}-\frac{z^2}{q+c^2}}}{\sqrt{q+a^2}+\sqrt{q+b^2}+\sqrt{q+c^2}}\\
&\left[\frac{q}{2}\left(\frac{1}{q+a^2}+\frac{1}{q+b^2}+\frac{1}{q+c^2}\right)+\left(\frac{ax}{q+a^2}\right)^2+\left(\frac{by}{q+b^2}\right)^2
+\left(\frac{cz}{q+c^2}\right)^2\right]
\end{aligned}
\label{eqa22}
\end{equation}
\begin{multicols}{2}
So, the $y$ component of electric field in the laboratory frame is:
\end{multicols}
\begin{equation}
E_y=\frac{\gamma Q}{6\pi^2\epsilon_0}\int_0^\infty dq\frac{e^{-\frac{x^2}{q+a^2}-\frac{y^2}{q+b^2}-\frac{z^2}{\gamma^2(q+c^2)}}}{\sqrt{q+a^2}+\sqrt{q+b^2}+\sqrt{q+c^2}}\frac{\left(qb^2-q^2+2b^4\right)y-2b^2y^3}{\left(q+b^2\right)^3}
\label{eqa23}
\end{equation}
\begin{multicols}{2}

\end{multicols}

\vspace{-1mm}
\centerline{\rule{80mm}{0.1pt}}
\vspace{2mm}

\begin{multicols}{2}

\end{multicols}

\clearpage
\end{CJK*}
\end{document}